\documentclass[twocolumn,showpacs,preprintnumbers]{revtex4}
\usepackage{graphicx}
\usepackage{dcolumn}
\usepackage{bm}
\usepackage{amsmath}
\usepackage{hyperref}

\begin{document}

\title{Emergent electromagnetism in solids }
\author{Naoto Nagaosa$^{1,2}$}\email{nagaosa@ap.t.u-tokyo.ac.jp}
\author{Yoshinori Tokura$^{1,2}$}\email{tokura@ap.t.u-tokyo.ac.jp}
\affiliation{$^1$ Department of Applied Physics, University of Tokyo, 7-3-1, Hongo,
Bunkyo-ku, Tokyo 113-8656, Japan\\
$^2$Cross-Correlated Materials Research Group (CMRG), and Correlated
Electron Research Group (CERG), RIKEN-ASI, Wako, Saitama 351-0198, Japan\\
}
\date{\today }

\begin{abstract}
Electromagnetic field (EMF) is the most fundamental field  
in condensed-matter physics.  Interaction between electrons,
electron-ion interaction, and ion-ion interaction are all of 
the electromagnetic origin, while the other 3 fundamental 
forces, i.e., gravitational force, weak and strong interactions are
irrelevant in the energy/length scales of condensed-matter 
physics.  Also the physical properties of condensed-matters 
such as transport, optical, magnetic and dielectric properties, 
are almost described as their electromagnetic responses.   
In addition to this EMF in the low energy sector, it often happens that
the gauge fields appear as the {\it emergent} phenomenon 
due to the projection of the electronic wavefunctions onto the curved manifold of the Hilbert sub-space. 
These {\it emergent electromagnetic fields} (EEMF's) play important roles in many places 
in condensed-matter physics including the quantum Hall effect,
strongly correlated electrons, and also in non-interacting electron systems.
In this article, we describe its fundamental idea and some of the 
applications recently studied. 
\end{abstract}

\maketitle

\section{Introduction and basic concepts}

Electromagnetism is the first gauge theory which was recognized in 
physics, and essentially contains the principle of the 
relativity. Most of the physical properties of solids 
are regarded as the responses of the systems to 
the electromagnetic field (EMF).  For example,
the transport phenomena are described by the currents 
induced by the electric field, magnetism is the behavior of
the magnetization in response to the magnetic field,
and optical properties are described by the electric polarization 
driven by the EMF of light.
On the side of the electronic system,  
the electron-electron and/or electron-ion interactions 
mediated by the EMF, i.e., mostly the Coulomb interaction, 
determine the quantum state even without the external  
EMF as a probe. This interaction often leads to the 
{\it emergent electromagnetic fields} (EEMF's) as will be described in more details below.
The other aspect of the electromagnetism in solids is the various 
{\it cross correlation effect}. Namely various different combinations of 
the inputs and responses have been explored. For example, an applied magnetic field
can induce the electric polarization, or the magnetization emerges 
under the electric field in the magneto-electric (ME) effect. 
This effect was proposed long time ago~\cite{Curie}, but has been very weak.
An explosion of the research was triggered by the discovery of the
multiferroic materials showing both the magnetic order and ferroelectricity 
which are coupled strongly to each other~\cite{KimuraNature,ReviewMF}.
These phenomena are 
driven by the internal degrees of freedom of electrons, i.e.,
spin, charge and orbital, which are again described coherently 
by the EEMF's.

 This EEMF is analogous to the usual EMF, but is
much richer because of the following aspects~\cite{Fradkin,Altland}.
(i) It is defined on the lattice because of the crystal structure 
and Bloch waves in solids, and the lattice gauge theories are relevant.
Therefore, various topological defects play important roles. 
(ii) One can generalize the space where EEMF is defined including the real space, 
momentum (${\bf k}$-)space, and other parameter space. 
(iii) The gauge group is not only U(1) as in the case of EMF, but can be more general 
such as the SU(2) corresponding to the Kramer's doublet in the time-reversal symmetric system, 
and U(N) corresponding to the N-bands system.   
(iv) Topological terms such as Chern-Simons term, WZW term, and $\theta$-term 
often appear to the effective action for EEMF.  Therefore, solids can be 
an ideal arena to study various concepts and techniques in quantum field theory~\cite{Peshkin}.

The basic reason why the EEMF's appear in condensed-matter is the
restriction of the Hilbert space. Namely, only the small portion of the
Hilbert space is relevant to the low energy phenomena, which constitutes
the fiber-bundle structure. For example, consider a vector moving on 
the surface of unit sphere $S^2$.  The parallel transport of this vector 
on $S^2$ leads to the change in its direction as shown in Fig.1. 
This is because $S^2$ has the (constant) curvature and the 
vector picks up this curvature through the connection 
for infinitesimal parallel transport. This connection corresponds to the 
vector potential of EMF while the curvature to the magnetic flux, which 
are related via Stokes theorem to each other. 
An analogy can be drawn between this geometry problem and the
quantum mechanics~\cite{Wilczek}. The vector corresponds to the wavefunction and
$S^2$ to the restricted Hilbert space. Therefore, the parallel transport 
on the subspace leads naturally to the gauge structure.  
We discuss below briefly the Berry phase~\cite{Berry} 
and the spin-orbit interaction from this viewpoint.

\subsection{Berry phase of electrons in perfect crystal}
  
The Bragg diffraction of the electrons by the periodic electric field of the ions 
in a crystal leads to the Bloch states as described by
\begin{equation}
\psi_{n {\bf k} \sigma} ({\bf r}, s) = e^{i {\bf k} \cdot {\bf r} }  u_{n {\bf k}} ({\bf r}) \chi_\sigma(s), 
\end{equation}
where $ u_{n {\bf k}}({\bf r}) $ is the periodic function with respect to the 
translation of lattice vectors, $\chi_\sigma(s)$ is the spin wavefunction 
for $\sigma= \uparrow, \downarrow$ with the spin coordinate $s= \pm$.
The band index $n$ and the crystal momentum ${\bf k}$ are the good quantum 
numbers and the energy eigenvalues $\varepsilon_n({\bf k})$'s with 
different $n$'s are usually separated by the band gaps. 
This means that the electrons in a solid are confined to each band in the lower energy
sector than the band gaps, which corresponds to the projection of the wavefunctions
to the sub-space of the Hilbert space.  This sub-space is usually "curved" and is
represented by the gauge connection. (This is an application of the 
Berry connection to Bloch electrons~\cite{Berry}.)
It is given by the expression
\begin{equation}
a_{n a} ({\bf k}) = -i \langle u_{n {\bf k}} | \partial_{k_a} | u_{n {\bf k}} \rangle
\label{eq:Berry}
\end{equation}
with $a=x,y,z$.
This quantity appears in the inner-product of the 2 wavefunctions
for the 2 neighboring ${\bf k}$-points as
\begin{equation}
\langle u_{n {\bf k}} |  u_{n {\bf k}+ \Delta {\bf k}} \rangle = \exp( i {\bf a}_{n} ({\bf k}) \cdot \Delta {\bf k} ).
\end{equation} 
This Berry connection leads to the concept of Berry curvature corresponding to
the magnetic field in the momentum space as
\begin{equation}
b_{n a} ({\bf k}) = \varepsilon_{a b c}  \partial_{k_b} a_{n c} ({\bf k}) 
\end{equation}
with the totally antisymmetric tensor  $\varepsilon_{a b c } $.

What is the physical consequences of this Berry phase ?
To answer to this question, it is useful to consider the motion of 
a wave-packet made of the Bloch states, where the position and momentum of a 
particle is defined within the accuracy consistent with the uncertainty 
principle~\cite{Sundaram}. In the presence of the Berry connection, the position operator ${\bf x}$  
originally defined as $x_\mu = i\partial / \partial k_\mu$ should be 
generalized to the gauge covariant form 
$x_\mu = i\partial / \partial k_\mu + a_{n \mu}(k)$. 
This is dual to the momentum operator $\pi_\mu$ in the presence of the 
magnetic field, i.e., $\pi_\mu = p_\mu + e A_\mu ( {\bf x} ) =- i \partial/\partial x_\mu + e A_\mu ( {\bf x} ) $, 
where $A_\mu({\bf x})$ is the vector potential of the EMF. 
Analogously to the commutator $[ \pi_x, \pi_y ] = ie B_z({\bf x})$ with $B_z({\bf x})$ 
being the $z$-component of the magnetic field, one can derive $[x,y] = i b_z({\bf k})$.
Namely, the real space coordinates do not commute with 
each other. 
With this commutator, the equation of motion for the wavepacket reads
\cite{Blount,Sundaram}
\begin{eqnarray}
{ { d x_\mu} \over { d t}} &=& -i[ x_\mu, H],
\nonumber \\
{ { d \pi_\mu} \over { d t}} &=& -i[ \pi_\mu, H].
\label{eq:EOM}
\end{eqnarray}
Here the Hamiltonian $H$ is given by $H= \varepsilon_n({\bf k}) + V({\bf x}) $ with
$\varepsilon_n({\bf k})$ being the energy dispersion of the band $n$ of interest, 
and $V( {\bf x} )$ is the slowly varying external potential. 
Putting this into eqs.(\ref{eq:EOM}) and we obtain
\begin{eqnarray}
{ { d x_\mu} \over { d t}} &=& 
{{ \partial \varepsilon_n({\bf k}) } \over {\partial k_\mu}}
 -i[ x_\mu, x_\nu] {{ \partial  V({\bf x}) } \over {\partial x_{\nu}}}
\nonumber \\
&=& {{ \partial \varepsilon_n({\bf k}) } \over {\partial k_\mu}}
+  \varepsilon_{\mu \nu \lambda} b_\lambda({\bf k}) 
{ { \partial  V ({\bf x}) } \over {\partial x_{\nu}} }
\nonumber \\
&=& {{ \partial \varepsilon_n({\bf k}) } \over {\partial k_\mu}}
+ ({\bf b} \times {\bf F})_\mu,
\nonumber \\
{ { d k_\mu} \over { d t}} &=& - {{ \partial  V({\bf x}) } 
\over {\partial x_{\mu}}}
= F_\mu,
\label{eq:EOM2}
\end{eqnarray}
where ${\bf F}= - \nabla_{{\bf x}} V ( {\bf x}) = e {\bf E}$ 
is the force acting on the electron.
In eq.(\ref{eq:EOM2}), the duality in the equation of motion is evident, i.e., 
the real space quantities $x_\mu, A_{\mu}({\bf x})$, $V ({\bf x})$
and those in the momentum space $k_\mu, a_{n \mu}({\bf k}), \varepsilon_n({\bf k})$
enter into the equations of motion in a symmetric way. Therefore one can
translate the phenomena in real (momentum) space to those in the 
momentum (real) space.  
Note, however, that compared with the
real space magnetic field which is divergence-free, i.e., $\nabla_{{\bf x}} \cdot {\bf B}(x)=0$, 
corresponding to the absence of the magnetic monopole in real space, 
$\nabla_{{\bf k}} \cdot {\bf b}({\bf k})$ can be non-zero in momentum space. 
Another important remark is that the symmetries give the following constraint.
The time-reversal symmetry $T$ gives the relation ${\bf b}_n ({\bf k}) = - {\bf b}_n (-{\bf k})$,
while the inversion symmetry $I$  ${\bf b}_n ({\bf k}) = {\bf b}_n (-{\bf k})$. Therefore, when
both $T$ and $I$ symmetries are there, there is no Berry curvature  ${\bf b}_n ({\bf k})$.
Also, in the noncentrosymmetric system where $I$-symmetry is absent,
${\bf b}_n ({\bf k})$ can be nonzero although the contributions from ${\bf k}$ and
$-{\bf k}$ cancel with each other. 
 
\subsection{Emergent SU(2) gauge field from Dirac equation}

Starting from the Dirac equation, the projection of 
the wavefunctions onto the low energy subspece leads 
to the gauge structure analogous to the electromagnetism.
This formalism is based on the real-space picture, and is 
complementary to the discussion in the last subsection
where the momentum space geometry has been considered.
When the SU(2) spin space is preserved, it leads to the
SU(2) non-Abelian gauge field which is coupled to the 
spin current, corresponding to the spin-orbit interaction (SOI).

We start with the quantum electrodynamics
(QED), where the Dirac relativistic electrons and their charge 
current is minimally coupled to EMF. Therefore, it appears 
that there is no chance for the spin current to take some
role in the electromagnetic phenomena. 
More explicitly, in the natural unit where 
$\hbar = c =1$, the Lagrangian of QED reads \cite{Peshkin}
\begin{equation}
L = {\bar \psi} [i \gamma^\mu {\hat D}_\mu - m ] \psi,
\end{equation}
where $\psi$ is the 4-component spinor field, 
${\bar \psi} = \psi^\dagger \gamma^0$,
$\gamma^\mu$ is the Dirac matrices, and 
${\hat D}_\mu = \partial_\mu - ie A_\mu$ 
is the gauge covariant derivative with $\mu=0,1,2,3$.
The 4-component charge current density is defined as
\begin{equation}
 j^\mu = - { {\partial  L} \over {\partial A_\mu} } = 
 -e {\bar \psi} \gamma^\mu \psi,
\end{equation}
whose 0-component is  the charge density $\rho$, 
while the spatial components are the current density ${\bf j}$. 
From the gauge invariance, the conservation law of the 
charge is derived through Noether's theorem as
\begin{equation}
\partial_\mu j^\mu = {{ \partial \rho} \over {\partial t} } + 
\nabla \cdot  {\bf j} = 0.
\end{equation}
By taking the variation, one can derive the Maxwell equation 
\begin{equation}
\partial_\mu F^{\mu \nu} = j^\nu,
\end{equation}
and Dirac equation
\begin{equation}
 [i \gamma^\mu {\hat D}_\mu - m ] \psi = 0.
\end{equation}
As is well-known, the solutions to Dirac equation 
are classified into 2 classes, i.e., the positive
energy and negative energy states separated by the 
twice of the rest
 mass energy of the electrons $2mc^2$.
Since the energy $mc^2$ is of the order of $MeV$,
for the low energy phenomena typically of the order of $ \sim eV$,
the negative energy states are not relevant. Therefore, the non-relativistic
Schroedinger equation is usually used which describes the 
dynamics of the 2-component spinor wavefunctions for positive energy states.
However, one needs to take into account one important 
aspect of "projection". Namely, the negligence of the negative energy
states means the projection of the wavefunctions onto the 
positive energy states, i.e, the sub-Hilbert space. 
Usually the sub-space is not flat but curved, and associated 
geometrical structure is introduced. The derivation of the 
effective Lagrangian describing the low energy physics is achieved by 
the expansion with respect to $1/(mc^2)$, and the result reads ~\cite{Froelich,He,Zaanen}
\begin{equation}
 L = i \psi^\dagger D_0 \psi   + \psi^\dagger  { { {\bf D}^2} \over { 2m}}  \psi +
{ 1 \over {2m}} \psi^\dagger \biggl[
eq \tau^a {\bf A} \cdot {\bf A}^a + { {q^2} \over 4} {\bf A}^a \cdot 
{\bf A}^a \biggr] \psi,
\label{eq:Lag}
\end{equation}
where $\psi$ is now the 2-component spinor and 
$D_0 = \partial_0 + ie A_0 + iq A_0^a { {\tau^a} \over 2}$,
and $D_i = \partial_i - ie A_i - iq A_i^a { {\tau^a} \over 2}$ $(i=1,2,3)$
are the gauge covariant derivatives with $q$ being the quantity
proportional to the Bohr magneton~\cite{Froelich,Zaanen}. 
$A_\mu$ is the vector potential for EMF, while the SU(2) gauge
potential are defined as 
$A_0^a = B_a$, $A_i^a =\epsilon_{ia \ell} E_{\ell}$. The former is coupled to 
the charge current, and the latter to the 4-component spin current 
$j^a_0= \psi \sigma^a \psi$, 
$j^a_i =  { 1 \over { 2 m i}}
[ \psi^\dagger \sigma^a D_i  \psi - D_i \psi^\dagger \sigma^a \psi ]$.
Note that the spin current is the tensor quantity with one suffix for the
direction of the spin polarization while the other for the direction of the flow.
Note an important difference between the EMF and the 
SU(2) gauge field. The former has the gauge symmetry, i.e., the freedom 
to chose the arbitrary gauge for the vector potential $A_\mu$,
while the "vector potential" $A^a_\mu$ for the latter 
is given by the physical field strength ${\bf B}$ and ${\bf E}$.
Actually, the relation $\partial^\mu A^a_\mu=0$ holds.    
Therefore, the SU(2) gauge symmetry is absent.
This is the basic reason why the 
spin is not conserved in the presence of the relativistic SOI. 
Instead, the spin current is "covariantly" conserved and  
satisfies~\cite{Froelich,He,Zaanen}
\begin{equation}
D_0 J^a_0 + {\bf D} \cdot {\bf J}^a = 0.
\end{equation}
This means that in the co-moving frame
the spin is conserved, while in the laboratory 
frame the spin source or sink appears when the 
electron forms a loop and comes back to the 
same position in space since the frame already changes~\cite{Zaanen}.
(Note that the usual SU(2) gauge theory is a nonlinear theory and 
the gauge field is "charged", and the sum of the 
spin current by matter field and the gauge field is conserved 
in the non-Abelian gauge theory as Yang-Mills first showed~\cite{Peshkin}. 
However, the SU(2) gauge invariance is absent here, and 
also the action $Tr[F_{\mu \nu}^2]$. There is no 
spin current from the non-Abelian gauge field when it is assumed 
to be the frozen background field.)

When the magnetic ordering occurs, the wavefunctions are further 
projected onto the spin component at each site, which is 
described by the field operator $\psi_\sigma$ of electrons 
decomposed into  
\begin{equation}
\psi_\sigma ({\bf x}) = z_\sigma ({\bf x}) f ({\bf x})
\label{eq:slave}
\end{equation}
with $f$ being the spinless fermion corresponding to the
charge degrees of freedom while $z_\sigma$ being the
2-component spinor corresponding to the
direction of the magnetization at ${\bf x}$. 
Putting eq.(\ref{eq:slave}) into eq.(\ref{eq:Lag}), we obtain the 
effective Lagrangian for $f$-field as
\begin{equation}
 L_{\rm eff.} = \psi^\dagger \biggl[i \partial_0 +  a^B_0 + a^{SO}_0 + A_0
  + { {( {\bf \nabla}  + i {\bf a}^B  + i {\bf a}^{SO} + i e {\bf A})^2}\over{2m}} 
\biggr] \psi, 
\label{eq:Leff}
\end{equation}
where 
$a^B_{\mu} = i \langle z | \partial_\mu |z \rangle$ is the U(1) field originating from the 
Berry connection of the spin wavefunctions, and 
$a^{SO}_{\alpha} =  A^a_\alpha \langle z | \tau^a  \partial_\alpha |z \rangle $,
$a^{SO}_0 =  A^a_0 \langle z | \tau^a  |z \rangle $ are U(1) field originating from the SOI.

Here a remark about the geometrical meaning of
the gauge field $a^B_{\mu}$ is in order~\cite{NagaosaLeePRB}. The magnetic flux 
made from ${\bf a}^B$, i.e.,   ${\bf b} = \nabla \times {\bf a}^B$, corresponds to 
the solid angle subtended by the spins. Namely, the scalar spin chirality 
given by ${\bf S}_i \cdot ({\bf S}_j \times {\bf S}_k) $ is reduced to 
$\bf{b}$ along the direction normal to the plane made by the
3 sites $i$, $j$, $k$ in the continuum limit. Therefore, the 
effective magnetic field is produced when the spin structure is
non-coplanar.  In summary, there are three "electromagnetic fields" in magnetic systems, 
i.e., (i) $a^B$ due to the Berry phase associated with the non-coplanar 
spin structure, (ii) $a^{SO}$ from the spin-orbit interaction (SOI), and 
(iii) usual Maxwell EMF $A$.

\subsection{Emergent electromagnetic field in correlated electronic systems}

Up to now, we have considered the gauge fields in the band structure or
magnetically ordered state, i.e., the single-particle properties,
where the gauge fields are the static background fields. 
The electron correlation effect corresponds to the 
fluctuating spin beyond the mean field theory, and accordingly the
gauge fields become dynamical~\cite{Polyakov}.
This issue has been studied extensively in the 
research of high temperature superconductors~\cite{NagaosaLeePRB,LeeNagaosaWen}.
As the simplest example, let us consider the non-linear sigma model 
describing the low energy physics of the quantum antiferromagnet~\cite{Fradkin,Wen}.
\begin{equation}
S = { 1 \over g} \int^\beta_0 d \tau \int d r |\partial_\mu {\bf n} |^2,
\label{eq:NLS}
\end{equation}
where the unit vector ${\bf n}$ specifies the staggered moment,
and $g \propto 1/S$ ($S:$ spin quantum number) 
is the dimensionless coupling constant 
representing the quantum fluctuation.
We introduce the spinor field $z = ^t(z_{\uparrow}, z_{\downarrow})$
with the constraint $\sum_\sigma z^\dagger_\sigma z_\sigma =1$,
${\bf n}$ can be represented by 
${\bf n} = z^\dagger_\sigma ({\bf \sigma})_{\sigma, \sigma'} z_{\sigma'}$
with ${\bf \sigma} = (\sigma_x,\sigma_y,\sigma_z)$ are the Pauli matrices.
Then $S$ in eq.(\ref{eq:NLS}) can be written as
\begin{equation}
S = { 1 \over g} | (\partial_\mu + i a_\mu )z_\sigma |^2
\label{eq:CP1}
\end{equation}
with the gauge field $a_\mu = -i z^\dagger_\sigma \partial_\mu z_\sigma$
corresponds to the Berry phase connection between the 2 spinors 
at the neighboring 2 spatial points.
Therefore, this $a_\mu$ corresponds to the U(1) gauge field
due to the spin chirality discussed in the previous subsection.
When we regard $z_\sigma$ as being the classical field, i.e., the 
condensed component corresponding to the magnetic ordering,
the gauge field $a_\mu$ is reduced to $a^B$ discussed in the 
last subsection. Without the magnetic ordering, on the other hand,
there is no condensation of $z$-field and hence the 
Higgs phenomenon for the gauge field $a_\mu$ does not occur.
In this spin liquid state, this gauge field is the dynamical 
degrees of freedom, i.e., obtains the effective action by integrating
over the rapid components of the spin fluctuations~\cite{Polyakov}.
(Note that there is no term describing the
 dynamics of the gauge field at the starting 
since it is introduced to express the constraint.)  
Therefore the gauge field appears dynamically as 
{\it emergent} phenomenon.    
The quantum spin fluctuations and electron correlation
effects are the subjects of intensive studies in condensed-matter
physics, and the gauge theoretical approach has been 
a powerful tool~\cite{LeeNagaosaWen,Fradkin,Altland}. 
Historically, motivated by the studies on strongly correlated
systems and quantum Hall systems from the gauge theoretical
viewpoint during 1990's, the conventional materials,
which can be well described by band theory and/or the mean field theory,
have been revisited from this new perspective. Below, we take some of
the examples, where new physical effects have been explored 
in {\it conventional} systems from the viewpoint of gauge fields.

\section{Materials and phenomena of emergent electromagnetism}

As described above, there are several sources of the 
EEMF, and correspondingly there are so many
related phenomena  in condensed-matters.
The common feature of these phenomena is that
the topological currents play an essential role.
In solids, there are many imperfections such as 
the defects and impurities, which cause the 
scattering of electrons and dissipation. This dissipation, i.e.,
Joule heating, is balanced with the energy supplied from the 
external electric field. This is the usual Ohmic current.
In the superconducting state, the macroscopic 
quantum coherence and the associated "rigidity" 
against the phase twist by the external magnetic field
produces the superconducting current.
This current does not decay nor cause any dissipation 
since the current-flowing state is in the
thermal equilibrium, and is separated by the macroscopic energy barrier
from the zero current state. 
To this list we would add here the third category of the current, i.e., the topological 
current, induced by the gauge field or curvature of the 
Hilbert space, i.e., EEMF. This third category includes 
the quantum Hall current and polarization current in ferroelectric 
materials, which are related to the Berry phase.
This topological current does not require the
off-diagonal long range order (ODLRO), and can exist 
in the normal states even at room temperature. 
Therefore, the topological currents will be of vital 
importance when the applications to electronics  
are considered. We will describe below several novel 
phenomena driven by this topological current in solids.
  
\subsection{Multiferroics}  

The minimal coupling to the SU(2) gauge field $A_\mu^a$ in eq.(\ref{eq:Lag})
or in eq.(\ref{eq:Leff}) for magnets means the close relation between the 
spin current $j_\mu^a$ and the electric polarization ${\bf P}$.
More explicitly, ${\bf P}$ is given by the derivative of the Lagrangian with respect to 
${\bf E}$, i.e.,
\begin{equation}
P_{i} \propto \epsilon_{i a \nu} j^a_\nu,
\label{eq:pspin}
\end{equation}
which means that the spin current produces the 
ferroelectric moment.  Then the next problem is how one can produce the
spin current. In the magnetically ordered state, the expectation value of
the spin current $ \langle  j_\mu^a \rangle \rangle$ in the ground state can be nonzero
for the noncollinear spin configuration. Consider the expression of the spin current
operator in the tight-binding model  
\begin{equation}
  j_\mu^a = i \sum_{i} t_{i i +\mu} c^\dagger_{i \alpha} (\sigma^a)_{\alpha \beta} c_{i+\mu \beta}   + h.c. ,
\end{equation}
where $t_{ij}$ is the transfer integral, $\sigma^a$ $(a=x,y,z)$ the Pauli matrices, and 
$\alpha, \beta$ the spin indices. For simplicity, let us consider the 
case of the spin ordering within the $xy$-plane, and the corresponding wavefunciton
$\chi_i =^t(  { 1 \over {\sqrt{2}} },   { { e^{i \phi_i} }  \over {\sqrt{2}} } )$.
Taking the expectation value with this wavefunction assuming the half-filling, one obtains
\begin{equation}
  \langle j_\mu^z \rangle \propto  {{t^2_{i i +\mu}} \over U}  \sin ( \phi_i - \phi_{i+\mu} ),
\end{equation}
where we employed the perturbation theory in $t/U$ ($U$: on-site 
Coulomb interaction or the Hund's coupling).
This is analogous to the Josephson current in superconductor with $\phi$ being the
phase of the order parameter, and the spin supercurrent in magnet is induced by the 
tilted spins. Putting this into eq.(\ref{eq:pspin}), 
$\langle P_z \rangle \propto  \epsilon_{z \mu \nu} \sin ( \phi_i - \phi_{i+\nu} )$.
More general expression is easily obtained for the polarization $P_{ij}$ obtained 
by the 2 spins ${\bf S}_i$ and ${\bf S}_j$ as
\begin{equation}
{\bf P} = \eta {\bf e}_{ij} \times ({\bf S}_i \times {\bf S}_j )
\end{equation}
with $\eta$ being the coupling constant related to SOI, 
and ${\bf e}_{ij}$ is the unit vector connecting the 2 sites
$i$ and $j$~\cite{KNB1,NNReview}. The vector product of the 2 spins  
${\bf \chi}_{ij} = {\bf S}_i \times {\bf S}_j $ is called vector
spin chirality. This quantity is even with respect to the
time-reversal operation $T$ similarly to the spin current. 

This generic argument is applied to the real materials as follows.
Compared with the free electrons in vacuum, 
the strength of the relativistic spin-orbit interaction can be enhanced by the factor of 
$\sim 10^6$ which is the ratio of the rest mass of the electrons $mc^2$ to the 
band gap. For 3d electrons in transition metal atoms, 
the SOI $\lambda$ is typically of the order of $\sim 20-50$meV, 
while it becomes $ \sim 0.5$eV for 5d electrons.
The electron correlation energy, on the other hand, decreases from 3d to 5d since
the wavefunction is more and more expanded.
In the cubic crystal field in transition metal oxides, 
the 5-fold degeneracy of d-orbitals is lifted due to the ligand field 
of oxygens. As the result, 3-fold degenerate $t_{2g}$ orbitals ($xy,yz,zx$-orbitals) 
with lower energy, and doubly degenerate  $e_g$-orbitals ($x^2-y^2, 3z^2-r^2$-orbitals) with 
higher energy are formed. 
The matrix elements of the orbital angular momentum ${\bf \ell}$ are zero
within the $e_g$-orbitals. On the other hand, they are nonzero among  $t_{2g}$-orbitals
and also between the $e_g$- and $t_{2g}$-orbitals. This SOI is the origin of the 
relativistic coupling between the magnetism and electric polarization.
To give a more explicit prediction for transition metal oxides, 
the cluster model of magnetic ions sandwiching an oxygen ion has been 
studied theoretically by taking into account the
spin-orbit interaction when deriving the super-exchange interaction~\cite{KNB1}. 
As mentioned above, the spin current flows 
between the 2 non-collinear spins ${\bf S}_i$ and ${\bf S}_j$,
which produces the electric polarization ${\bf P}$ as given by  
\begin{equation}
{\bf P} \cong  -{ {4e} \over 9} {( {V \over \Delta})}^3 I {\bf e}_{ij}
\times( {\bf S}_i \times {\bf S}_j ),
\label{sxP}
\end{equation}
where $I=\langle p_x|z|d_{zx} \rangle $, and 
$\Delta$ ($V$) is the energy difference (hybridization) between the p-orbitals and the d-orbitals.
The SOI interaction is implicitly included in this model by picking up one doublet
after splitting by the SOI. 
Applying this result to various magnetic structures, one can easily 
predict the presence or absence, and the direction of the polarization.
This theory is not contradicting with the symmetry argument developed 
for magnets \cite{Landau,Harris06}, but stresses the physical mechanism of 
the spin current-induced polarization. 

A recent experimental breakthrough is the discovery of 
the multiferroic behavior in $R$MnO$_3$($R$=Gd,Tb,Dy)~\cite{KimuraNature,ReviewMF}. 
In these materials, the spontaneous electric polarization $P_s$
appears accompanied by the magnetic order, and they are necessarily strongly
coupled. 
It has been also revealed that the magnetic structure 
which induces the electric polarization is  the cycloidal spiral 
in good accordance with the theoretical 
prediction above~\cite{Kenzelmann,yamasaki1,Yamasaki08}.
Figure 4 shows the representative multiferroic 
behavior of DyMnO$_3$. In this material, the spin rotation plane 
flop from bc- to ab-planes, and accordingly the direction of the
electric polarization changes from c- to a-axes~\cite{ReviewMF}. 

Now the extensive experimental studies have been done 
to explore the multiferroic materials, e.g., Ni$_3$V$_2$O$_8$~\cite{NVO},
Ba$_{0.5}$Sr$_{1.5}$Zn$_2$Fe$_{12}$O$_{22}$~\cite{BSZFO},
CoCr$_2$O$_4$~\cite{CoCr2O4}, MnWO$_4$~\cite{MnWO4},
CuFeO$_2$\cite{CFO}, LiCuVO$_4$~\cite{LiCuVO4}, and
LiCu$_2$O$_2$~\cite{LiCu2O2} are discovered to be 
multiferroics. Namely, multiferroicity is not a special
phenomenon but is a rather ubiquitous phenomenon in 
Mott insulators. These experimental findings urged 
the systematic theoretical studies on the microscopic 
mechanisms of the spin-related electric polarization~\cite{han1,han2}. 
The perturbative approach in both $V/\Delta$ and 
$\lambda/\Delta$ is employed, where $V$ and $\Delta$ represent the transfer integral 
and the charge transfer energy between the transition-metal (TM) $d$ and 
ligand (L) $p$ orbitals. 
This analysis concludes that the polarization 
${\bf P}_{{\bf r}+{{\bf e} \over 2}}$ appearing on the bond between the 
sites $\bf{r}$ and $\bf{r}+\bf{e}$ is given by
\begin{eqnarray}
\bf{P}_{\bf{r}+{ \bf{e} \over 2}} &=& P^{\rm{ms}}
(\bf{m}_{ \bf{r}}\cdot \bf{m}_{ \bf{r} +\bf{e}}) \bf{e}
+ P^{\rm{sp}}
\bf{e} \times (\bf{m}_{ \bf{r}}\times \bf{m}_{\bf{r} + \bf{e}}) 
\nonumber \\
&+& P^{\rm{orb}}\left[ (\bf{e}\cdot\bf{m}_{\bf{r}})
\bf{m}_{\bf{r}}-
(\bf{e}\cdot\bf{m}_{\bf{r}+\bf{e}}) \bf{m}_{\bf{r}+\bf{e}}\right],
\label{eq:mech}
\end{eqnarray}
where $\bf{m}_{\bf{r}}$ is the spin direction at $\bf{r}$.  
The first term $P^{\rm{ms}} \propto (V/\Delta)^3$ is the 
polarization due to the exchange-striction,
which is nonzero when the inversion symmetry between 
$\bf{r}$ and $\bf{r}+\bf{e}$ is absent because the
2 intermediate states of doubly occupied d-orbitals
becomes inequivalent. This term does not requires the 
SOI, hence is considered to be 
larger than the rest of the terms if it exits.
The second term $P^{\rm{sp}}\propto(\lambda/\Delta)(V/\Delta)^3$ is 
due to the spin current mechanism already discussed. 
The third term  
$P^{\mathrm{orb}}\sim \mathrm{min}(\lambda/V, 1) (V/\Delta)$
is nonzero for the partially filled $t_{2g}$ orbitals 
and comes from the modification of the single-spin
anisotropy due to the electric field~\cite{han1,han2}.
These three contributions appear differently depending on the 
wavevector of the polarization. Now the origins of 
the multiferroic behavior of various materials are investigated 
and are classified according to these 3 mechanisms in most
of the cases~\cite{ReviewMF}.

  In eq.(\ref{eq:mech}), the polarization is due to the 
relativistic SOI (spin current mechanism or spin anisotropy mechanism) or the broken inversion symmetry
at the center of the bond (exchange-striction).
Therefore, it is an important question what is the most ubiquitous and generic interaction between the 
electric field and the spin system without these mechanisms. 
By a careful analysis of the super-exchange processes in Mott insulators 
taking into account the quantum dynamics of the spins in the intermediate states, 
we have derived the following effective Lagrangian~\cite{MostovoyNomura} 
\begin{equation}
L_E = - \int d^3 {\bf r} T_{ab} E_a e_b,
\end{equation}
where ${\bf E}$ is the electric field and 
\begin{equation}
{\bf e}=  {1 \over 2} \sin \theta ( \partial_t \theta \nabla \varphi - \partial_t \varphi \nabla \theta)
\end{equation}
is the electric field component of EEMF associated with the spin structure derived from the 
U(1) Berry connection. 
Here ${\bf n} = ( \cos \varphi \sin \theta, \sin \varphi \sin \theta, \cos \theta)$ is the 
unit vector of the magnetization in the continuum approximation, and
$T_{ab}$ is the tensor which depends on the details of the atomic configurations.
One important remark here is that $T_{ab}$ vanishes in the single-orbital case
since the spins form singlet and has no quantum dynamics in the intermediate
states in the processes of the super-exchange interaction. 
However, in most of the cases, the orbital degrees of freedom is
active and $T_{ab}$ is nonzero. 
This coupling is analogous to the spin motive force in the metallic 
ferromagnetic systems by which the domain wall or vortex motions produce the voltage 
drop~\cite{Berger,Maekawa,Niu}.

\subsection{Topological Hall effects}
As explained in eq.(\ref{eq:EOM2}), the most natural phenomena expected from the 
gauge fields in momentum space are related to the transverse motion of 
the electrons to the external electric field, i.e., the Hall effects.
Contrary to the usual Hall effect driven by the Lorentz force, 
we call the Hall effects originating from the gauge fields as 
{\it topological Hall effects}, and we describe below some of the 
examples. 

\subsubsection{Spin-orbit interaction and anomalous Hall effect}

Hall discovered the Hall effect in metallic ferromagnetic systems
due to the spontaneous magnetization instead of the external 
magnetic field~\cite{AHE}. This effect is called anomalous Hall effect (AHE)
and its mechanism has been a controversial issue for more than 
a century. It is agreed that the effect is due to the 
SOI combined with the spontaneous magnetization, but 
the issue was the role of the impurity scatterings. 
Intrinsic mechanism was first proposed by Karplus-Luttinger (KL)~\cite{KL}, 
who considered the anomalous velocity for the first time 
discussed in section I. 
However, their theory has been 
criticized by Smit~\cite{Smit} saying that the impurity scattering,
which is inevitable and also indispensable to reach the steady 
state under electric field, invalidates the KL theory. 
This posed the question if the
dissipationless topological current can survive even in the
presence of the dissipative Ohmic current. 
Smit~\cite{Smit} proposed the skew scattering mechanism, where the
impurity scattering in the presence of SOI gives the 
asymmetry of the transition rates 
between $W_{{\bf k} \to {\bf k'}}$ and $W_{{\bf k'} \to {\bf k}}$.
This breakdown of the detailed balance leads to the
net current perpendicular to the applied electric field,
i.e., the Ohmic transport current is slightly distorted 
in the perpendicular direction. Later Berger~\cite{SJ} 
proposed another mechanism called side-jump, where the 
transverse shift of the electron trajectory occurs
at the scattering in the presence of the SOI. These 2 mechanisms are
called extrinsic ones as opposed to the intrinsic one by KL.  
  The recent advances in this problem is two-folds~\cite{AHE}.
One is the recognition that the band structures of the 
ferromagnetic materials can be topologically nontrivial 
characterized by Chern numbers. 
Haldane~\cite{Haldane} was the first to notice that the quantum Hall effect can 
be realized without the Landau level formation in a tight-binding model  
on a lattice with the complex transfer integrals. What is recognized recently 
is that this scenario can be realized in the ferromagnetic materials
with the SOI~\cite{MOnoda}. In this case, even though the overlaps of the band dispersions
make the system metallic, one can define the Chern number of each band 
when the gap opens at each ${\bf k}$-point. Therefore, the 
ferromagnetic metal could be an {\it implicit} quantum Hall system
with $k_z$-dependent Chern numbers.    
The other is the state-of-art first-principles calculations of electronic 
states and the Hall response as well as the 
accurate experimental measurements of the
AHE in various materials~\cite{AHE}.   
 In short, the transverse conductivity $\sigma_{xy}$ due to
the intrinsic mechanism can be written 
in terms of the Berry phase curvature, i.e., 
the magnetic flux $b_{nz}({\bf k})$ for the 
Bloch wave state with crystal momentum ${\bf k}$ and the 
band index $n$~\cite{TKNN}:
\begin{equation}
\sigma_{xy} = {{e^2} \over { 2 \pi \hbar}}  \sum_{n {\bf k}} 
f(\varepsilon_n({\bf k})) 
b_{nz}({\bf k}),
\label{eq:TKNN}
\end{equation}
where the $f(\varepsilon_n({\bf k}))$ is the Fermi distribution function 
for the energy $\varepsilon_n({\bf k})$ of the Bloch state, and 
the vector potential $a_{n \mu} ({\bf k})$ is defined in eq.(\ref{eq:Berry}),
and ${\bf b}_{n}({\bf k}) = \nabla_k \times {\bf a}_n({\bf k})$. 
The meaning of this expression is that the Hall current
is the sum of the anomalous velocities of Bloch states occupied by electrons in 
the equilibrium distribution. 
Therefore the transverse conductivity $\sigma_{xy}$ represents the gauge 
field distribution in the momentum space. Especially when the chemical 
potential is in the gap, the ${\bf k}$-integral in eq.(\ref{eq:TKNN}) is over the whole 
first Brillouin zone (1st BZ), and one might think it is zero due to the periodicity 
with respect to ${\bf k} \to {\bf k}+ {\bf G}$ with ${\bf G}$ being the reciprocal lattice vector. 
Namely the contour integral over boundary of the 1st BZ appears to cancel. 
However this is not always the case, and the integral is 
$(e^2/h) \times {\rm integer}$ especially in 
the 2-dimensions.  This integer is the 
topological number called Chern number for the $U(1)$ fiber bundle of the 
Bloch wavefunction \cite{TKNN}. 
 The finite Chen number means that the phase of the 
Bloch wavefunction can not be defined continuously over 
the whole 1st BZ analogous to the Yang-Wu construction of  Dirac monopole~\cite{Dirac,Fradkin,Altland}. 
Here the plural patches need to be introduced to cover the  1st BZ.
The relevance of the Dirac monopole to the Hall effect 
is understood more explicitly in the following 3-dimensional model.
\begin{equation}
H({\bf k}) = \sum_{a = x,y,z} k_a \sigma_a.
\end{equation}
The corresponding magnetic field associated with the Berry phase reads
\begin{equation}
{\bf b}_{\pm}({\bf k}) = \pm { {{\bf k}} \over { 2 |{\bf k}|^3} }, 
\end{equation}
where $\pm$ corresponds to the 2 bands with energy 
$\varepsilon_{\pm} ({\bf k})= \pm |{\bf k}|$. 
This means that the magnetic monopole exists
at the band crossing point ${\bf k}={\bf 0}$, 
and $\nabla_k \cdot b_{\pm}({\bf k})= \pm 2 \pi \delta( {\bf k})$  
has the delta-functional singularity. 
The contribution from the fixed $k_z$ is given by 
\begin{equation}
\sigma_{xy}(k_z) = \pm {{e^2} \over { 2 h}} {\rm sign} {k_z}. 
\end{equation} 
For the 2-dimensional system where $k_z$ is regarded as 
a parameter $m$ characterizing the system, it describes the
quantum Hall system with $\sigma_{xy} = \pm { {e^2} \over {2 h}} $.
The factor $1/2$ is not realized in real system because 
the Dirac fermion appears always in the pair, i.e., 
species doubling, in the 1st BZ and hence the 
Chern number is an integer. 

  Essential to AHE is the fact that the nontrivial topological structure 
described above is not special as in the case of quantum Hall system, 
but is generic for magnets in the presence of the SOI. A simple 
2-dimensional 3-band model for $t_{2g}$ orbitals was constructed, 
and the uniform magnetization produces the nonzero Chern number for each 
band~\cite{MOnoda}. In the absence of the SOI, the 
up and down spin bands are independent of each other except the 
exchange energy splitting, and there occurs several
band crossing points but no finite Chern numbers.
Then SOI lifts the degeneracies
to generate the mass term $m$ as discussed above and 
produce the finite Chern numbers. This is the reason why the SOI 
can not be treated perturbatively in sharp contrast to the 
previous theoretical treatments. 
This nontrivial behavior has been found both in the first-principles
band calculation and in the experiment in SrRuO$_3$, 
a metallic ferromagnet with $T_c=130$K \cite{Fang}. 
In Figs. 5 shown the transport properties of SrRuO$_3$. 
Figure 5(c) shows the temperature dependence of the Hall resistivity which is 
dominated by the anomalous contribution. Just below $T_c$, it is negative
and turns into positive and again has the maximum.
This characteristic behavior strongly suggests that the perturbative 
expansion in $\lambda M$ ($\lambda$: SOI, 
$M$: magnetization) is not allowed and the AHE is a fingerprint of the 
Berry phase of the Bloch wavefunction.
We plot in Fig.5(d) the transverse conductivity $\sigma_{xy}$ as a function 
of the spontaneous magnetization compared with the first-principles band
structure calculation. This non-monotonous temperature dependence
including the sign change is due to the band crossings acting  
as the magnetic monopoles in momentum space. 
Accordingly Fig. 6 shows the distribution of the Berry curvature $b_{n z}({\bf k})$ as a 
function of $(k_x,k_y)$ at fixed $k_z=0$. 
It shows a sharp peak around $k_x=k_y=0$ and the
ridges along $k_x = \pm k_y$. This sharp peak represents the monopole 
corresponding to the band crossing. 
When the Fermi energy $\varepsilon_F$
is near the magnetic monopoles, the electrons are subject to the strong gauge 
field and hence the contribution to the Hall constant is resonantly enhanced. 
The integral over $k_z$ leads to the partial cancellation of the 
positive and negative contributions, but still the rapid change
of $\sigma_{xy}$ as a function of $\varepsilon_F$ results. This explains 
both the first-principle band calculation and experimental results, which 
are shown in Figs.5(c) and (d). 
Similar conclusion is obtained also for the AHE in Fe \cite{Fe},
where the sharp spiky structures of the Berry phase distribution 
has been found.

An important recent development is the experimental 
measurements of {\it dynamical} AHE, i.e.,
$\sigma_{xy}(\omega)$ in ferromagnetic materials.
Thanks to the rapid progress of the telahertz spectroscopy~\cite{Iguchi,Shimano},
the low frequency structures of the dynamical AHE are being revealed.
Especially, the dominance of the contributions from the  band crossings
is expected to appear in the $\omega$-dependence of  $\sigma_{xy}(\omega)$
as well as the non-monotonous temperature dependence of $\sigma_{xy}(\omega=0)$.
Figure 7 shows an example of the low frequency data for AHE in SrRuO$_3$. 
It is clearly seen that nontrivial structure exists at around 4meV, which is 
well fitted by a model assuming the band crossing near the Fermi energy.

Up to now, it is argued that in some materials, the AHE can be explained 
by the intrinsic mechanism based on the Berry phase concept,
but the controversy between the intrinsic and extrinsic needs to 
be resolved. As for this problem, again the role of the enhanced 
Berry curvature near the band crossings and the topological nature
are crucial.  In ref.\cite{OSN}  a model with a band anti-crossing is considered and
its Hall response is calculated taking into account the disorder scatterings.
The Hall conductivity $\sigma_{xy}$ is obtained as a function of 
$\sigma_{xx}$ representing the strength of the disorder. 
Figure 8 summarizes the results including the various experimental data
plotted. (Note that the calculation has been done for a 2D model, and 
$\sigma$'s are scaled by the lattice constant along the $z$-direction 
which is assumed to be $\cong 0.4$nm.)  This result captures the
respective role of the transport and topological currents. 
In the very clean region where $\sigma_{xx} > 10^6 \Omega^{-1} cm^{-1}$,
the skew scattering mechanism (extrinsic mechanism) is dominant 
and hence $\sigma_{xy} \propto \sigma_{xx}$.  
When the disorder strength becomes larger, the 
transport current is suppressed much more 
rapidly compared with the topological current, the
latter of which is "protected" topologically,
leading to the plateau region in the region 
 $10^4 \Omega^{-1} cm^{-1}< \sigma_{xx} < 10^6 \Omega^{-1} cm^{-1}$.
 In this intermediate region, to which many of the metallic
ferromagnets belong, the intrinsic mechanism is  dominant and
hence the first-principles calculations can predict the AHE 
semiquantitatively. When the disorder becomes even stronger,
$\sigma_{xy}$ begins to decrease obeying the approximate
scaling law  $\sigma_{xy} \propto \sigma_{xx}^{1.6}$.
This scaling is in good accordance with the experimental 
results as seen in Fig. 8, but its understanding is still left
for future studies. To conclude this subsection, AHE offers an interesting laboratory
where  both the Ohmic transport and topological currents 
coexist, and it has been established that the latter can 
contribute to the Hall effect even in the metallic systems.
Quantized version of the AHE, where most of the bulk states
are localized~\cite{QAHE} or the chemical potential is in the gap~\cite{QAHE2}, have been 
discussed also. In this case, the current is carried by the edge
channels as in the case of quantum Hall system.

\subsubsection{Scalar spin chirality and anomalous Hall effect} 

  There are a vast variety of the non-collinear spin structures 
found experimentally. These systems are the ideal 
laboratory to test the idea of U(1) EEMF discussed in section I.
Of particular interest is the idea of "scalar spin chirality"
$\chi_{ijk}$ defined for the 3 spins as
\begin{equation}
\chi_{ijk} = {\bf S}_i \cdot ( {\bf S}_j \times {\bf S}_k ),
\end{equation}
which is $T$-odd. 
As discussed in section IB and Fig.2, the scalar spin chirality acts as
an effective magnetic field and leads to the Hall response~\cite{Ye,Ohgushi}.
First, we consider the case where the magnetic unit cell is small, and 
the Bloch wavefunctions are defined in the 1st BZ. 
The total gauge flux penetrating the unit cell is zero or integer multiple of
$2 \pi$ due to the periodicity since the contour 
integral of ${\bf a} (r)$ along its boundary vanishes ( mod $2 \pi$) 
due to the periodicity.
On the other hand, the gauge field corresponding to the Berry 
phase becomes nonzero for the Bloch wavefunctions in
the momentum space with a possible finite Chern number of each band.  
Roughly 
speaking, this happens when the unit cell contains the more than 2 
different types of the loop, and each band feels the fluxes with 
different weight to obtain the Chern number.
Ohgushi et al. \cite{Ohgushi} showed that this idea can be
realized in a ferromagnet on Kagome lattice 
with the non-coplanar spin configuration.

  Pyrochlore lattice can be regarded as the 3-dimensional generalization 
of the Kagome lattice, i.e., pyrochlore lattice contains the 
Kagome lattice normal to the [111] direction. In the material 
Nd$_2$Mo$_2$O$_7$ (hereafter we denote it as NMO), there are 2 
interpenetrating sublattices composed of the tetrahedrons of Nd and Mo atoms 
shifted along the c-axis as shown in Fig.9(a)~\cite{TaguchiScience}. 
The conduction electrons are on the Mo sublattice, while the
localized spins on Nd sublattice are subject to the strong 
anisotropy to form non-coplanar spin configurations. 
The localized spins and conduction electron spins are coupled 
antiferromagnetically, which transmit the 
scalar spin chirality of Nd to conduction electrons (Fig.9(b)).
Figure 9(c) shows the temperature dependence of the
Hall resistivity which shows the  steep increase
as the temperature is lowered corresponding to the 
increase of the exchange field from Nd spins and 
associated scalar spin chirality.
The absolute value of the low temperature value is
consistent with the theoretical calculation taking into 
account the tilting angle of the Mo spins estimated from the
neutron scattering experiment~\cite{TaguchiScience}.
Furthermore,  the observation of
the sign change of $\sigma_{xy}$ under the external magnetic field
along [111] direction \cite{TaguchiPRL} is consistent with
the sign change of the spin chirality. 

 The analysis in ${\bf k}$-space above is justified when the 
magnetic unit cell is small compared with the mean free path $\ell$.
In this case, the EEMF is defined by the Berry phase in 
${\bf k}$-space. In the other limit, i.e., when $\ell$ is shorter
than the size of the spin texture $\xi$, the EEMF 
is defined more appropriately in ${\bf r}$-space, which acts
as the effective EMF with $\omega_c \tau <<1$
($\omega_c$: cyclotron frequency, $\tau$: mean free time). 
This situation is realized in the recently discovered Skyrmion
crystals in noncentrosymmetric magnets with B20 structure~\cite{MnSi,Yu}.
MnSi, (Fe,Co)Si, and MnGe are the examples of this class
of materials, which has Dzyaloshinskii-Moriya (DM) interaction as
described by the following Hamiltonian in the continuum approximation.
\begin{equation}
H = \int d {\bf r} \biggl[
{ J \over 2} (\nabla {\bf M}({\bf r}) )^2 + \alpha {\bf M}({\bf r})  \cdot (\nabla \times {\bf M}({\bf r}))
- {\bf H} \cdot {\bf M}({\bf r})
\biggr],
\label{eq:DM}
\end{equation}
where ${\bf M}({\bf r})$ is the magnetization, ${\bf H}$ the magnetic field,
$J$ the ferromagnetic interaction, and
$\alpha$ the DM interaction constant. The lattice constant is taken to be unity.  
The ground state under zero magnetic field is the single
spiral with the spins rotating in the plane perpendicular to the
wavevector ${\bf q}$. The length of $q= |{\bf q}|$ is determined to the
the ratio $\alpha/J$ while its direction is determined by the 
weak spin anisotropies not included in the
Hamiltonian eq.(\ref{eq:DM}).
   
Recently, a neutron scattering experiment identified the 
mysterious A-phase in MnSi as the Skyrmion crystal state stabilized by the 
external magnetic field and thermal fluctuations~\cite{MnSi}. 
The conical spin structure is the most stable state under the magnetic field
in the major part of the phase diagram in 3D crystal.
On the other hand, when one reduces the thickness of the sample smaller than the 
wavelength of the spiral,  the conical state becomes energetically higher than
the Skyrmion crystal when the external magnetic field is perpendicular to the film. 
Actually, a Monte Carlo simulation of the
2D magnet with DM interaction concluded that the Skyrmion crystal state is
stable in much wider region of the phase diagram in $(T,B)$-plane 
($T$: temperature, $B$: magnetic field ) including the zero temperature
case\cite{Yi}. Motivated by this study, a recent experiment using the 
Lorentz tunneling electron microscopy (TEM) succeeded in real-space observation of the Skyrmion 
crystal in the thin film of (Fe,Co)Si~\cite{Yu}. 
Figure 10 shows the phase diagram and the Lorentz TEM images of the 
Skyrmions and Skyrmion crystals. The left panel shows the 
experimental results, while the right panel the results by Monte Carlo
simulations, which indicate the good agreement. 

As shown in Fig. 11 schematically, each Skyrmion wrap the sphere once in 
the order parameter space of ${\bf M}$, which means that the integral 
of the {\it magnetic} flux ${\bf b} \parallel {\hat z}$ over one Skyrmion is $2 \pi$. From this 
one can estimate the effective magnetic field induced by the 
Skyrmion. It is typically 4000T when the size of the Skyrmion $\xi=1$nm, and 
inversely proportional to the square of $\xi^2$.
Since MnSi, (Fe,Co)Si, MnGe, are metallic systems,
the conduction electrons are coupled to the spins and spin chirality.
Therefore, the Hall effect is expected due to the Lorentz force 
driven by  ${\bf b}$ replacing the external magnetic field ${\bf B}$.
Recently, the Hall measurement on MnGe has been done 
and the contribution from the Skyrmion has been analyzed~\cite{Kanazawa}.
The Hall resistivity $\Delta \rho_{yx} \cong 200$n$\Omega$cm is due to this
topological contribution, which is corresponding to the effective magnetic
field $b \cong 1100T$ for $\xi= 3$nm.  
These values are compared with the case of MnSi, where
$\xi= 18$nm, $b \cong 28T$, and $\Delta \rho_{yx} \cong 5$n$\Omega$cm.

The dynamics of Skyrmions and Slyrmion crystal is an intriguing issue to 
be studied.  A recent experiment~\cite{Skmotion} shows the Skyrmion crystal motion driven by 
a current with the threshold value $j_c \cong 10^2$A/cm$^2$m, which is 
orders of magnitude smaller than that for the domain wall motion in 
ferromagnets~\cite{Ono}. Theoretically, the motion of the 
Skyrmions will induce the {\it electric field} ${\bf e}$ as 
electromotive force due to $d {\bf b}/dt$.  This leads to several
physical consequences such as a new mechanism for the 
damping of the Skyrmion motion, and the motion of 
the Skyrmion transverse to the current (Skyrmion Hall effect)~\cite{Jiadong}.

\subsubsection{Hall effect of light}

The topological Hall effect does not require that the particles are
charged, and hence is possible even for the uncharged particles. 
A representative example
of uncharged particle is the photon, and one can ask if the Hall effect 
of light can occur or not.
The constraint which induces the Berry curvature in this case 
is that the electric and magnetic fields of light is always perpendicular to the 
direction of the propagation,
i.e., that of the wavevector ${\bf k}$. This situation corresponds to the
strong coupling limit of SOI in electrons. 
Actually, the effect of the Berry phase in the propagation of light
in the optical fiber has been studied both experimentally and theoretically,
and the rotation of the polarization was confirmed for the light with the 
${\bf k}$-vector slowly changing and enclosing an area on the sphere~\cite{Wu}.

Note that the quantum nature of the photon is not required 
for this Berry phase effect, which is solely due to the wave
nature of light. Therefore, this offers a unique opportunity to 
study the Berry phase effect in classical systems.
Extending this idea, we have derived the semiclassical equation of
motion for the wavepacket of light including the effect
of the finite wavelength~\cite{HEL}. (Note that the wave-optics is reduced 
to the geometric optics in the short-wavelength limit, 
the former of which corresponds to
the quantum mechanics while the latter to the classical Newtonian mechanics.)
This semiclassical equations of motion for the position ${\bf r}$, the wavevector (momentum)
${\bf k}$ and the spin state $|z \rangle$ (2-component spinor describing the polarization of light) reads
\begin{eqnarray}
{ {d {\bf r}} \over { d t} } &=& 
v ({\bf r})  { {\bf k} \over { |{\bf k}|} } +  { {d {\bf k}} \over { d t} } \times \langle z|{\bf \Omega}_{{\bf k}}|z\rangle,
\nonumber \\
{ {d {\bf k}} \over { d t} } &=& - [ \nabla v({\bf r}) ] {\bf k},
\nonumber \\
{ {d | z \rangle } \over { d t} } &=& - i  { {d {\bf k}} \over { d t} } \cdot \bf{\Lambda}_{{\bf k}} | z \rangle,
\label{eq:opt}
\end{eqnarray}
where $v({\bf r})= c/n({\bf r})$ is the velocity of light in the medium, 
$[{\bf \Lambda}_{{\bf k}}]_{\lambda \lambda'} = -i {\bf e}^\dagger_{\lambda {\bf k}} 
\nabla_{\bf{k}} \bf{e}_{\lambda' \bf{k}}$ is the connection of $2 \times 2$ matrix form, i.e., SU(2)
gauge connection, and ${\bf \Omega}_{{\bf k}} = \nabla_{{\bf k}} \times  {\bf \Lambda}_{{\bf k}} + 
{\bf \Lambda}_{{\bf k}} \times {\bf \Lambda}_{{\bf k}}$ is the field strength.
It is diagonal in the basis of the right and left-circular polarization, i.e.,
${\bf \Omega}_{{\bf k}} = \sigma_3 {\bf k}/ |k|^3$. 
This equation indicates that the Berry curvature has the opposite
sign for opposite circularly polarized light, which gives the 
polarization-dependent anomalous velocity 
adding to the group velocity of wavepacket. More explicitly,
one can consider the reflection and transmission of light
at the interface of 2 media with different dielectric constants~\cite{HEL}.
The anomalous velocity is finite when the particle is subject to
the gradient of the refractive index $\nabla n({\bf r})$, and results 
in the finite transverse shifts of the transmitted and reflected lights
at this reflection/transmission. These shits are in the opposite direction
for opposite circular polarization. This effect has recently been 
observed experimentally using the "weak measurement"~\cite{Ill}.

The idea of the Berry phase for light can be generalized to the
photonic crystal, where the Bloch waves are formed. 
Especially, one can consider the crystal with distortion, 
where the Berry phase is defined in the 6-dimensional space $({\bf k}, {\bf r})$. 
Especially, the curvature is enhanced near the gap edge when the gap is small.
This situation is realized for the X-ray propagating in crystals since the deviation of the 
dielectric function from unity is typically of the order of $10^{-6}$, and hence the
"periodic potential" for the Bloch wave is very weak. 
Sawada et al.~\cite{Sawada} considered the role of this enhanced Berry curvature 
in the propagation of X-ray in deformed crystal. 
The prediction is that the shift in the trajectory $\Delta {\bf r}$ of the
X-ray is given by
\begin{equation}
\Delta {\bf r} \cong {\bf G} ({\bf G} \cdot {\bf u} )  { {\omega} \over {\Delta \omega}} { 1 \over {|{\bf k}|^2} },
\end{equation}
where ${\bf G}$ is the reciprocal lattice vector satisfying the Bragg condition for
the wavevector ${\bf k}$ of the X-ray, ${\bf u}$ the displacement of the crystal, 
$\omega$ the frequency of X-ray, and $\Delta \omega$ is the gap in the Bloch wave of 
X-ray. As mentioned above, $\omega/\Delta \omega$ can be as large as $10^6$, which determines 
the magnification of $\Delta {\bf r}$ compared with ${\bf u}$ considering that
$|{\bf G}| \sim |{\bf k}|$.  This theoretical prediction has been recently confirmed experimentally 
using the single crystal of Si~\cite{Ishikawa}, and offers a new 
principle for the X-ray microscope. 

\subsubsection{Hall effect of magnons}

Another example of the uncharged particle is the magnons
in insulating magnets. Although the charge current is not available in this case, 
the energy current can be carried by the magnons and hence the 
thermal Hall effect without resorting to the Lorentz force  is 
When one considers the propagation of the magnon wave in a ferromagnet,
it behaves as a Bloch wave and the band structure is formed. 
When more than 2 atoms are in the unit cell, the Berry phase 
is generally expected for this Bloch wavefunction.
We consider the Hamiltonian with the Dzyaloshinskii-Moriya (DM) SOI as 
\begin{equation}
H = - \sum_{ij}  J_{ij} {\bf S}_i \cdot {\bf S}_j + {\bf D}_{ij} \cdot ({\bf S}_i \times {\bf S}_j )
- \sum_i {\bf H} \cdot {\bf S}_i.
\end{equation}
Suppose the external magnetic field ${\bf H}$ and 
ferromagnetic moments align along $+z$-direction. 
Then, the spin wave Hamiltonian can be written as
\begin{equation}
H_{SW} = - \sum_{ij} {\tilde J}_{ij}S ( e^{-i \phi_{ij} } b^\dagger_i b_j + h.c. )
+ \sum_i H^z b^\dagger_i b_i,
\end{equation}
where  ${\tilde J}_{ij} e^{i \phi_{ij} } = J_{ij} + i D^z_{ij}$.
This indicates that the DM interaction acts as a vector 
potential and effective magnetic flux for the propagating magnons~\cite{magnonhall}. 
Similar to the case of AHE, the distribution is such that
the total effective magnetic flux is zero (or equivalently the integer
multiple of $2 \pi$) when integrated over the unit cell. Therefore, 
finite effect survives when the inequivalent loops exist in the
unit cell. Kagome lattice and its 3D generalization pyrochlore lattice are the
representative crystal structures satisfying this condition.
And actually the thermal Hall effect has been recently observed 
in an insulating ferromagnet Lu$_2$V$_2$O$_7$ with pyrochlore 
structure~\cite{magnonhall}.
In the pyrochlore structure, the midpoint between any 2
corners of a tetrahedron is not an inversion symmetry center, 
and hence the nonzero DM interaction is expected.
Symmetry further determines the direction of the 
DM vectors ${\bf D}_{ij}$ as shown in Fig. 12(a). 

In this material, the spontaneous
magnetization $M$ emerges below Curie
temperature $T_c=70K$, which is isotropic and almost coincides
with 1 Bohr magneton ($\mu_B$) at low temperatures, indicating the
collinear ferromagnetic state with spin $S =1/2$. 
Below $T_c$, the thermal Hall conductivity $\kappa_H$ is
discernible,  whereas it is very small above 80 K.
The magnitude of the thermal Hall conductivity
has a maximum at around $50 K$. Similar to the
magnetization, the thermal Hall conductivity steeply
increases and saturates in the low magnetic field
region. Therefore, the observed thermal Hall effect
is due to the spontaneous magnetization $M$ 
as in the case of AHE. 
$\kappa_H$ gradually decreases with increasing the
magnetic field furthermore due to the opening
of the gap in the magnon spectrum 
induced by the magnetic field.
All those effects are contained in following theoretical expression
for  $\kappa_{\alpha \beta}$~\cite{magnonhall}:
\begin{eqnarray}
\kappa_{\alpha \beta} &=& \Phi_{\alpha \beta} 
{{ k_B^2 T} \over { pi^{3/2} \hbar a}} 
\biggl( 2 + { {g \mu_B H} \over { 2 J S} } \biggr)^2 
\nonumber \\
&\times&
\sqrt{ {{k_B T} \over { 2 J S}} } {\rm Li}_{5/2} 
\biggl[ \exp \biggl( -  { {g \mu_B H} \over { k_B T} } \biggr) \biggr],
\label{eq:kappa}
\end{eqnarray}
where $\Phi_{\alpha \beta} = - \varepsilon_{\alpha \beta \gamma} n_{\gamma} D/ ( 8 \sqrt{2} J)$
( $\varepsilon_{\alpha \beta \gamma}$: totally antisymmetric tensor, ${\bf n}$: direction of the magnetization,  
  $a$: lattice constant, $D = | {\bf D}_{ij} |$), and ${\rm Li}_n(z) = \sum_{k=1,\infty} z^k/k^n$.
 Figure 12(b) shows the comparison between the experimental result and eq.(\ref{eq:kappa}).
The only adjustable parameter is the ratio  $D/J = 0.32$,
which is a reasonable value for transition metal
oxides. 
From these facts, it is convincing that the thermal Hall 
effect observed in Lu$_2$V$_2$O$_7$ is due to the
magnons affected by the Berry curvature due to the
DM interaction.

\subsubsection{Spin Hall effect}

The concept of the anomalous velocity can be generalized to the
time-reversal $T-$symmetric systems. As already discussed in 
eq.(\ref{eq:opt}), the SU(2) Berry curvature is non-zero even in 
the $T-$invariant system, and the analogous effect is expected 
for the electrons also.
This idea has been explored for the band structure of 
semiconductors~\cite{MNZ}.
In GaAs and Ge, the valence bands consist of 
three p-ortbitals ($L=1$) while the conduction bands s-orbitals ($L=0$). 
Therefore the 6 bands (3 times the spin degeneracy 2) constitute the valence 
bands, which are split into 2-fold degenerate split-off (SO) bands ($J=1/2$), 
and the 4-fold degenerate bands ($J=3/2$) at the $\Gamma$-point ($k=0$). 
The 4-fold degenerate bands are further split into 2 doubly-degenerate 
heavy-hole and light-hole bands, i.e., 2 Kramers degenerate bands at finite ${\bf k}$. 
(Here we neglect the breaking of the inversion symmetry in the case of GaAs, 
which is small.) These 4 bands are described by Luttinger Hamiltonian as

\begin{eqnarray}
H_0 &=& \sum_{{\bf k}}   c_{\mu, {\bf k}}^{\dagger}
H_{\mu \nu}({\bf k}) c_{\nu, {\bf k}}, \nonumber \\
H_{\mu \nu}({\bf k}) &=& \frac{1}{2m}
\left( (\gamma_{1}+\frac{5}{2}\gamma_{2})k^{2}-
2\gamma_{2}({\bf \hat k} \cdot {\bf S} )^{2} \right)_{\mu\nu}.
\end{eqnarray}

As discussed above, the degeneracy acts as the magnetic monopole
for the Berry curvature, and the $\Gamma$-point is especially interesting since the 2 
doubly-degenerate bands touch there. Since there are Kramers degeneracy 
at each ${\bf k}$-point, the Berry connection is the $2 \times 2$ matrix ($SU(2)$). 
There are 2 pseudo-spin states, i.e., parallel and 
anti-parallel pseudo-spin to ${\bf \hat k} = {\bf k}/| {\bf k}|$, which is called helicity. 
The gauge field and hence the anomalous velocity depend on the helicity of the 
state, i.e., it has the opposite sign for different helicity states. 
Therefore, even though there is no net charge current, the external 
electric field can induce the transverse spin current. This is called
the (intrinsic) spin Hall effect, and for the case of
the p-doped GaAs/Ge the spin current $j^\mu_\alpha$ was 
predicted to be \cite{MNZ}
\begin{equation}
J_j^i = \sigma_H^s \epsilon^{ijk} E_k
\label{eq:SHE}
\end{equation}
with 
\begin{equation}
\sigma_H^s = { 1 \over {6 \pi^2}} (k_F^H - k_F^L),
\end{equation}
where $k_F^H(k_F^L)$ is the Fermi wavenumber for the heavy (light) holes. 
This intrinsic spin Hall effect is driven by the band structure and its
Berry phase connection, in sharp contrast to the 
extrinsic mechanisms previously proposed by Dyakonov and Perel 
in 1971 \cite{DP} and others \cite{Hirsch,SZhang}. 
Sinova et al. independently proposed the intrinsic spin Hall effect
for n-GaAs using the Rashba model~\cite{Sinova}. 
Note that eq.(\ref{eq:SHE}) looks natural when one remember that the
SU(2) gauge potential  $A_i^a =\epsilon_{ia \ell} E_{\ell}$ is coupled to the
spin current as discussed in section I. However, the 
topological structure of the Bloch wavefunction is needed to 
realize the intrinsic spin Hall effect.

There are several experimental reports on the spin Hall effect.
In semiconductors, the spin accumulation near the edge of the 
sample is measured to confirm the spin Hall effect. 
Kato et al. \cite{Kato} observed the spin accumulation 
at the edge of the n-type GaAs in terms of the Kerr rotation.
Wunderlich et al. \cite{Wunderlich} observed the circular polarized
LED from the recombination process of the holes in the 
p-type GaAs. Both experiments seems to be relevant to the 
spin Hall effect, but its origin, i.e., intrinsic or extrinsic,
is still controversial~\cite{ReviewSHE}. 
The experiment on metals usually measures the inverse
spin Hall effect, i.e., the voltage induced by the injection of
spin current from the ferromagnetic metals~\cite{metallicshe}. 
For some of the metals, the first-principles band structure
calculations are applied to predict the intrinsic spin Hall effect 
in e.g. Pt, which agrees reasonably well with the experimental 
observations~\cite{Guo}. The readers are referred to a 
review~\cite{ReviewSHE} for more details.

\section{Other interesting systems and conclusions}

There are several important and intriguing systems from the viewpoint
of EEMF which are not covered in section II.
One natural question to ask is "What is the global aspect of the sub-Hilbert 
space and gauge fields, and their implications on the 
physical properties ?". 
Of particular importance from this viewpoint is the topological insulators~\cite{TIReview}.
One can imagine the 2 copies of quantum Hall systems
with opposite chiralities for up and down spins. This systems will 
show the quantized spin Hall conductance due to the 
nonzero Chern numbers $C_{\uparrow} = - C_{\downarrow}$.
The spin Chern number is defined as  $C_{\uparrow} - C_{\downarrow}$,
which can be finite even for the $T$-symmetric systems.
However, in the presence of the SOI, usually all the components of the
spin are not conserved, and hence the spin current and Chern number 
are not well defined. Actually,  the spin Hall conductance 
in spin Hall insulator, where the chemical potential is
inside the band gap but still the spin Hall conductance is finite due
to the band inversion, is not quantized~\cite{volvik,SHI}. 
Kane-Mele discovered a $T$-symmetric model with SOI, which shows
a stable nontrivial topological phase with the helical edge channels even with the
mixing of the spins~\cite{TIReview}. This phase is characterized by the $Z_2$ topological number 
instead of the Chern number. This idea is now generalized to 3-dimensions 
and also to arbitrary dimensions to classify all the possible topologically nontrivial
states as far as the band structure can be defined~\cite{Schnyder}. 
The basic concept is that the Bloch wavefucntions as the fiber bundle
can have the globally nontrivial structure characterized by the 
index defined as the integral of the Berry curvature or EEMF over the 1st BZ.
Combining the mapping between the different 
dimensions and symmetry class enables the classification of
the topological classes including both the insulators and superconductors
with gap. Recently the generalization to include the ${\bf k}$- and ${\bf r}$-spaces has been 
achieved also~\cite{Teo}. The readers are referred to recent reviews for more details~\cite{TIReview}.

Noncentrosymmetric systems with SOI are also interesting 
laboratory to study the topological nature and EEMF. 
Rashba system is the representative one described by the Hamiltonian 
\begin{equation}
H_R = {{ {\bf p}^2} \over {2 m}} + \lambda {\bf e}_z \cdot ( {\bf s} \times {\bf p}),
\label{eq:Rashba}
\end{equation}
where ${\bf e}_z$ is the direction of the potential gradient which breaks $I$-symmetry, 
${\bf s}$ and ${\bf p}$ are the spin and momentum operators, respectively.
This interaction has been experimentally demonstrated for the 2-dimensional electrons 
at the interface of GaAs system~\cite{Nitta} or at the interface electrons in oxides system ~\cite{STO1,STO2},
and also the bulk 3-dimensional material BiTeI~\cite{Ishizaka}.
The spin splitting at each ${\bf k}$-point means that the spin and the velocity 
is tightly coupled, and hence the transport and magnetic properties.
Therefore, the various cross correlations between the magnetic 
and transport degrees of freedom are expected and actually observed 
experimentally in the Rashba systems, which constitute the important 
part of the spintronics. The spin Hall effect, spin Galvanic effect,
and dynamical magneto-electric effect are the examples of this cross-correlation
driven by the Rashba interaction~\cite{Chernyshov,Miron}.
Combined with the superconducting order parameter, the Rashba type SOI 
leads to even more interesting phenomena~\cite{Bauer}.  Because of the 
absence of the inversion symmetry, the classification of 
even and odd parity does not work there, and hence the 
singlet and triplet pairings are mixed. This fact leads 
to several unusual features such as the $H_{c2}$ beyond the 
Pauli limit~\cite{Bauer}. Furthermore, it is shown that the topological 
superconductor with helical Majorana edge channels as the Andreev bound states
can be realized when the p-wave pairing is dominant over the s-wave pairing component~\cite{Tanaka}.
When the magnetic field or exchange field is applied to open the gap 
at the band crossing and the Fermi energy is within that gap, 
the spinless pairing superconductor is realized, and the 
Majorana fermions are expected to appear there~\cite{TanakaReview}.  
 These are consistent with the more general classification scheme of the
topological superconductors~\cite{Schnyder}.   

In this article, we have discussed the physics of emergent electromagnetic 
field (EEMF) in condensed-matter systems. The gauge connections and fields are
naturally introduced in various situations when the wavefunctions are constrained
on some manifold in the Hilbert space. In the band structure and/or in the 
mean field theory, this gauge fields are not dynamical but frozen one or are 
controlled as parameters by the external electromagnetic field (EEM). 
However, the physical phenomena driven by EEMF are even richer 
than those of EMF because the lattice gauge theory, non-Abelian gauge fields, 
higher dimensions, topological terms, are often relevant to EEMF. 
In this sense, the gauge fields can be a guiding and unifying principle  
in condensed-matter physics. Needless to say, the dynamical nature of EEMF in correlated systems
and spin liquids, which was not covered in this article, also remains to be 
an important subject, and continue to be a central issue in the 
condensed-matter theory~\cite{Fradkin,Wen,Altland}.

The author thanks H. Katsura, M. Mochizuki, S. Murakami, N. Furukawa, A.V.Balatsky, M. Onoda, S. Onoda, H.J.Han, C. Jia, 
K. Nomura, M.Mostovoy, S.C. Zhang for collaborations, and T. Arima, N. Kida, M. Kawasaki,  
D. I. Khomskii, and A. Aharony for useful 
discussions. This work was supported by Priority Area Grants, 
Grant-in-Aids under the Grant numbers 19048015, 19048008, and
21244053, and NAREGI Nanoscience Project from the Ministry of
Education, Culture, Sports, Science, and Technology, Japan,
Strategic International Cooperative Program (Joint Research Type) 
from Japan Science and Technology Agency, 
and by Funding Program for World-Leading Innovative 
R and D on Science and Technology (FIRST Program)D

\clearpage
\begin{figure}[t]
\begin{center}
\includegraphics[width=10cm,angle=-90]{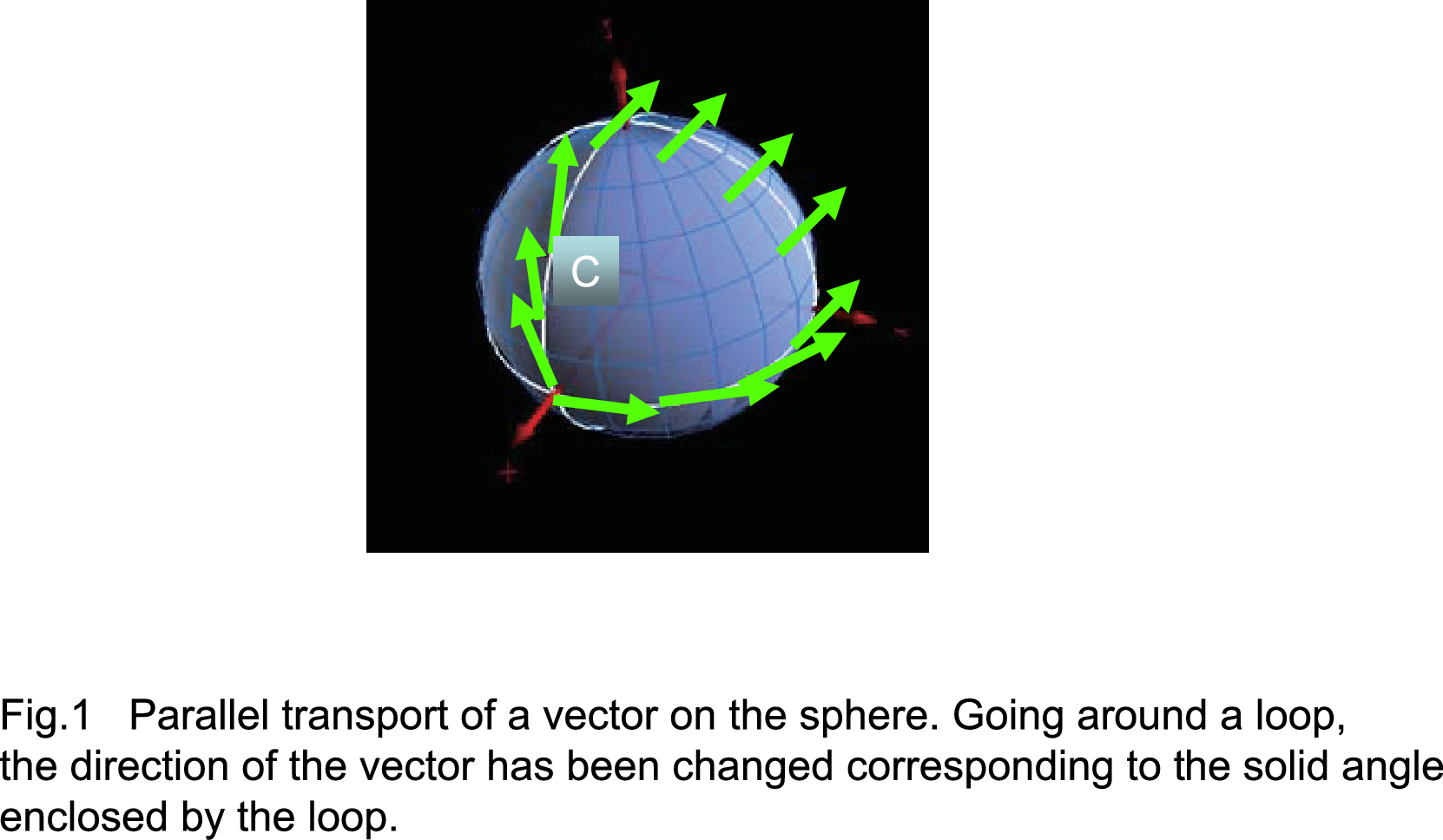}
\end{center}
\end{figure} 

\clearpage

\begin{figure}[t]
\begin{center}
\includegraphics[width=9cm,angle=-90]{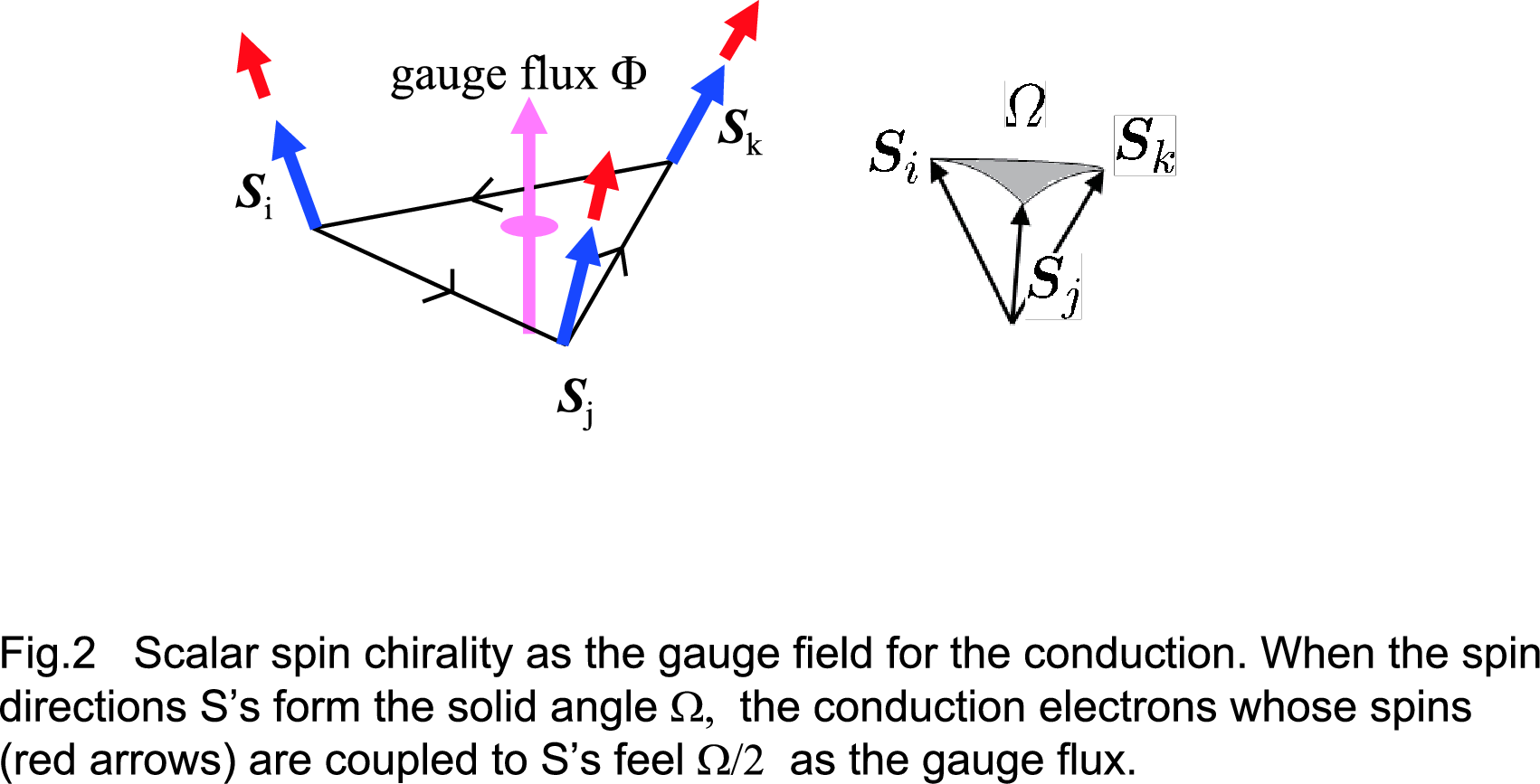}
\end{center}
\end{figure} 

\clearpage

\begin{figure}[t]
\begin{center}
\includegraphics[width=10cm,angle=-90]{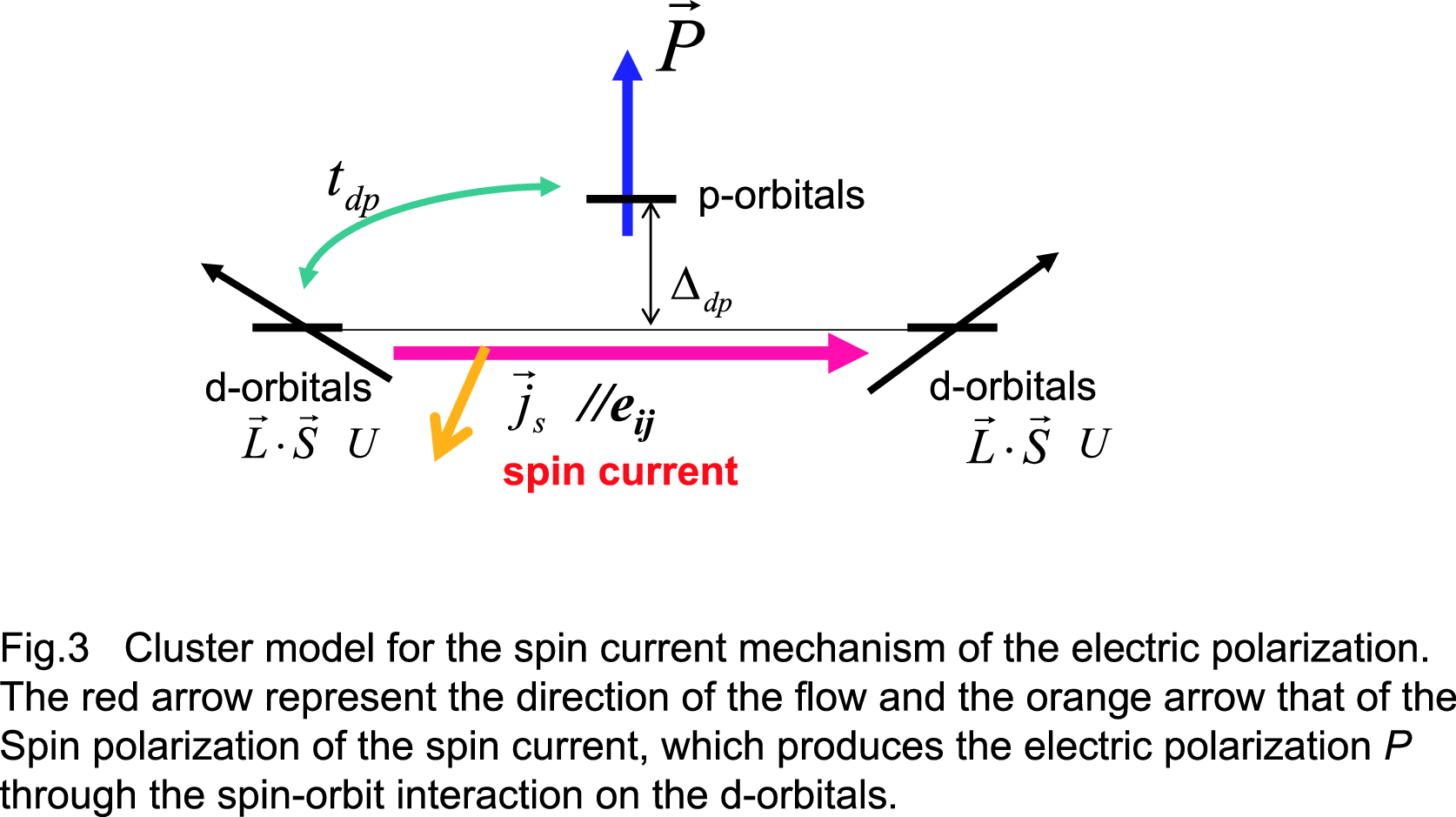}
\end{center}
\end{figure} 

\clearpage

\begin{figure}[t]
\begin{center}
\includegraphics[width=14cm,angle=-90]{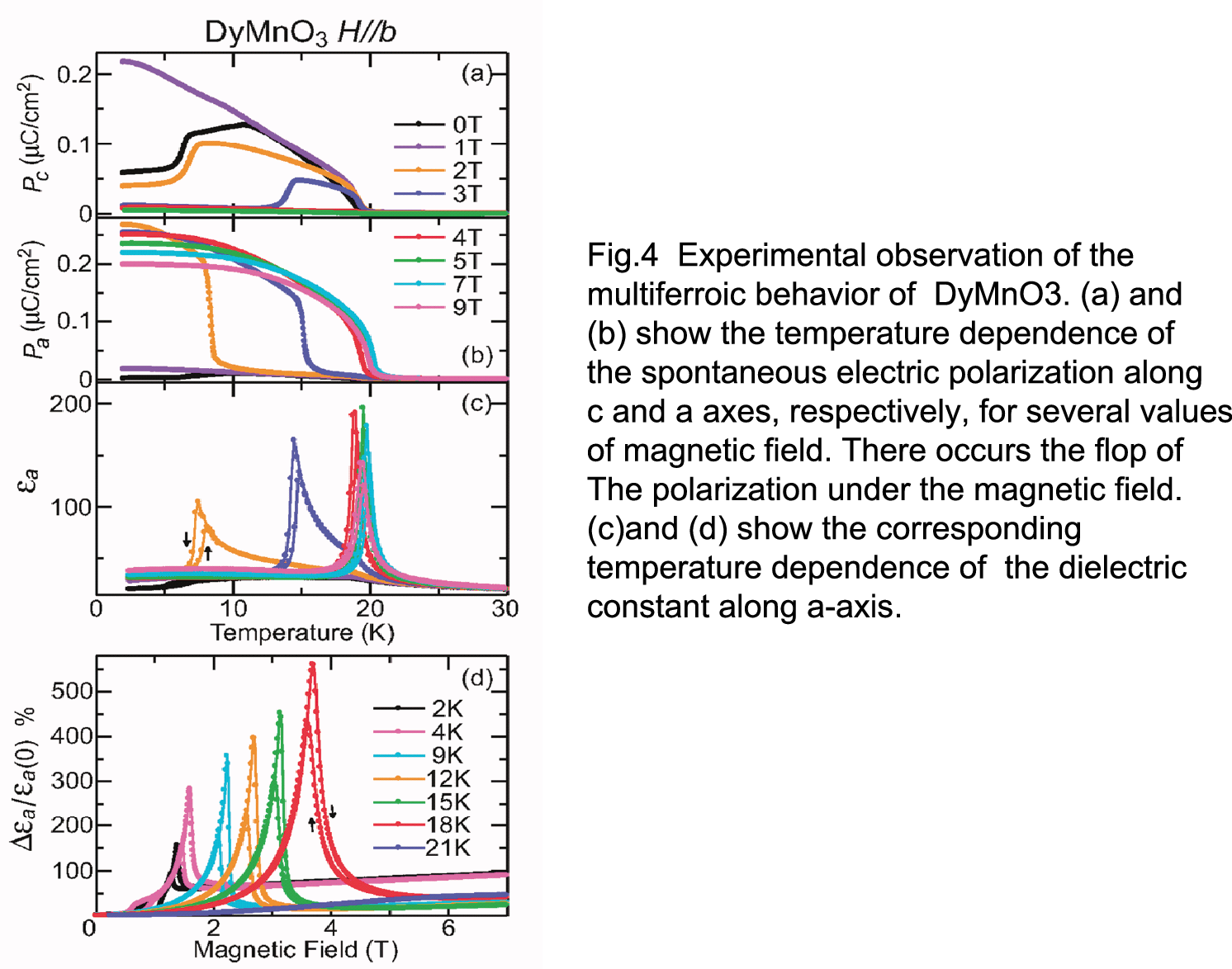}
\end{center}
\end{figure} 

\clearpage

\begin{figure}[t]
\begin{center}
\includegraphics[width=12cm,angle=-90]{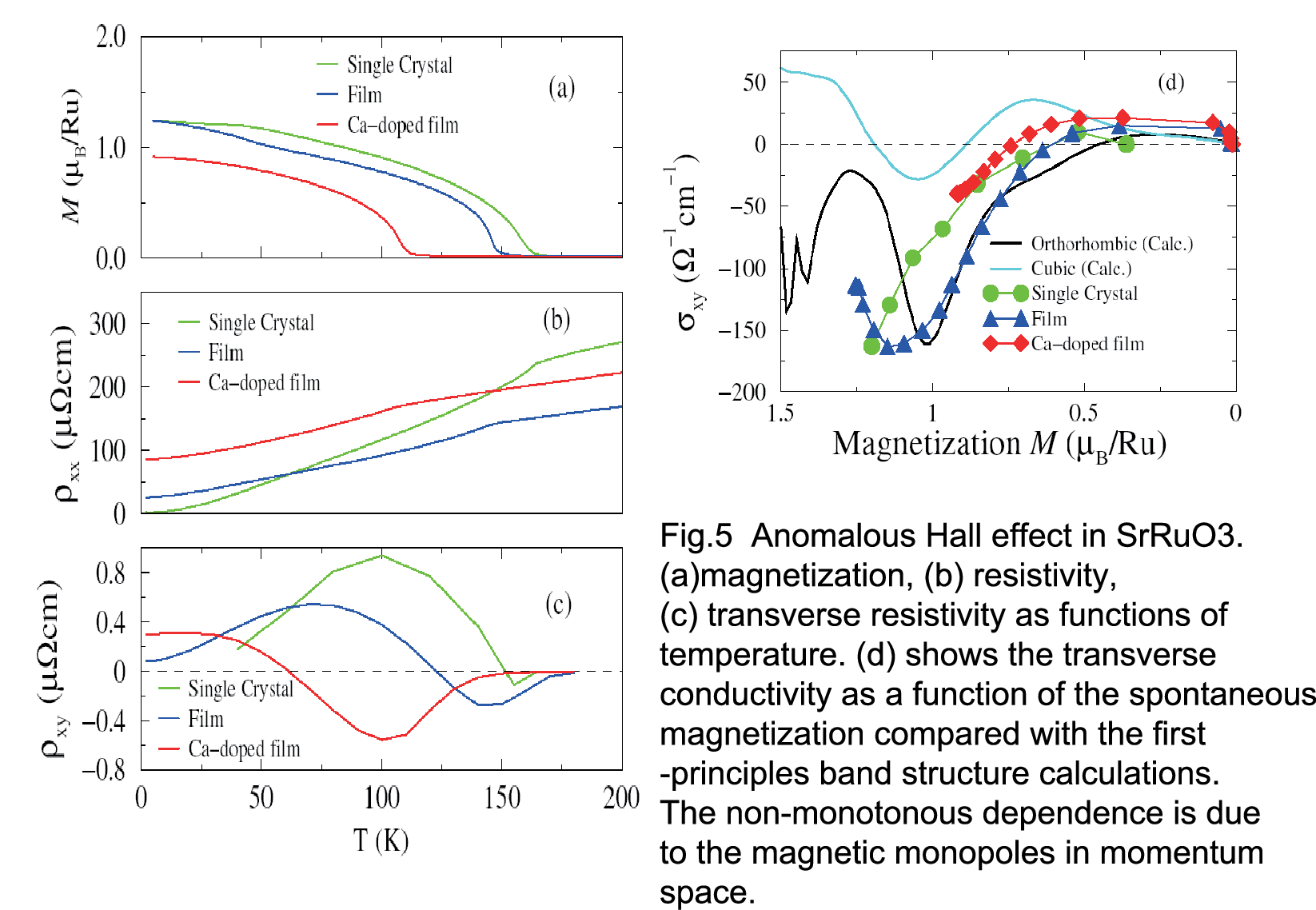}
\end{center}
\end{figure} 

\clearpage

\begin{figure}[t]
\begin{center}
\includegraphics[width=10cm,angle=-90]{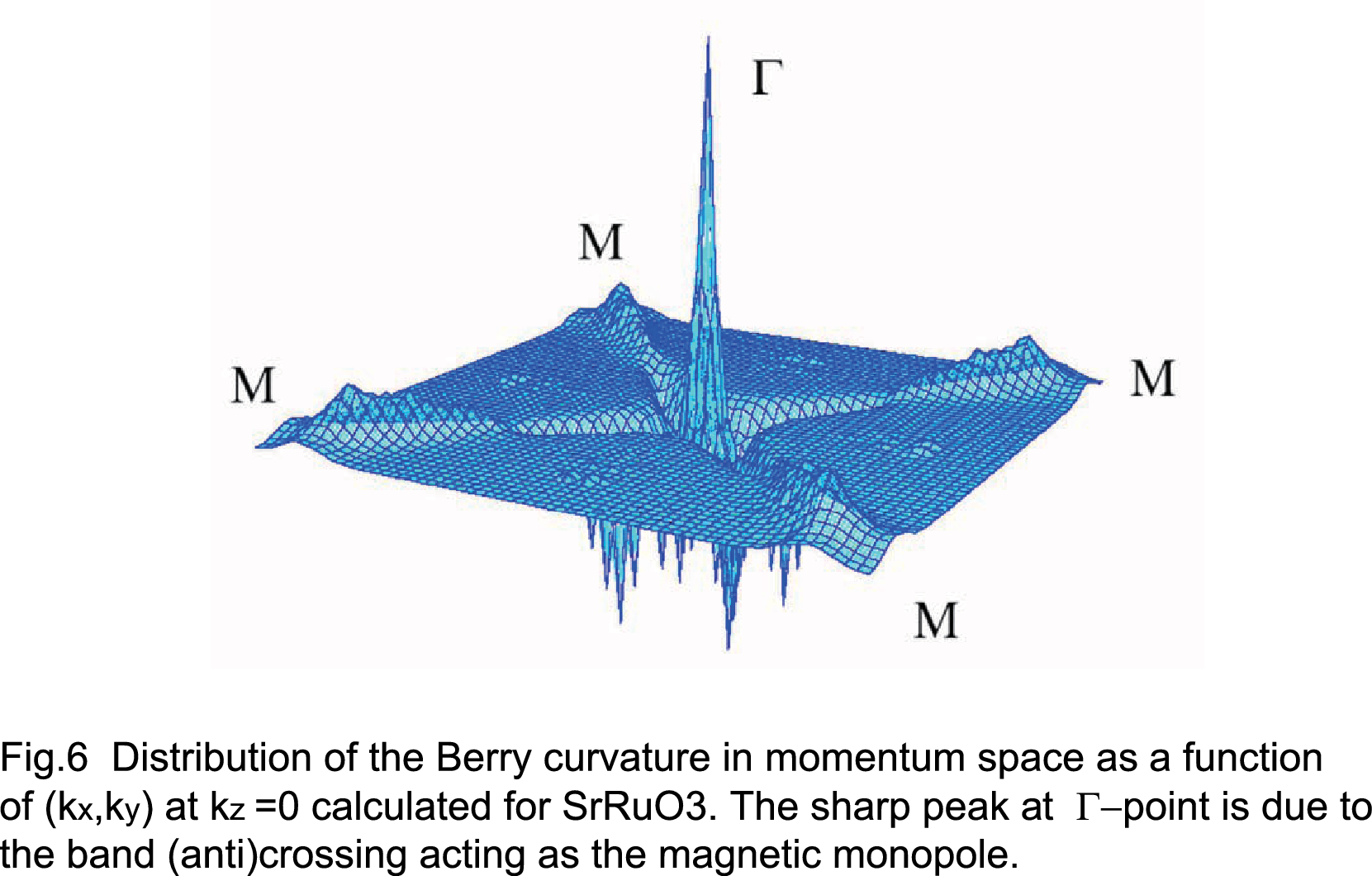}
\end{center}
\end{figure} 

\clearpage

\begin{figure}[t]
\begin{center}
\includegraphics[width=18cm]{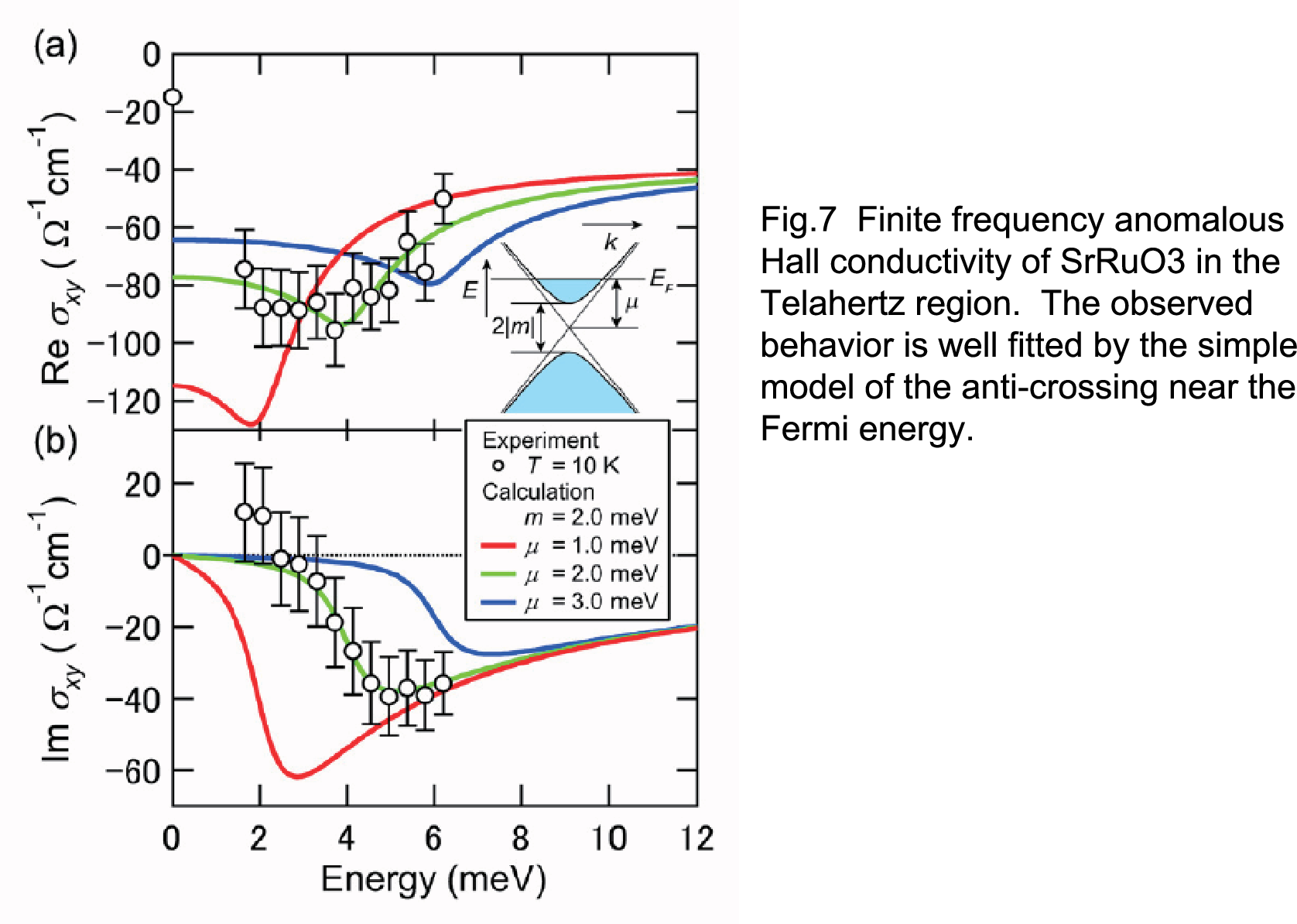}
\end{center}
\end{figure} 

\clearpage

\begin{figure}[t]
\begin{center}
\includegraphics[width=16cm]{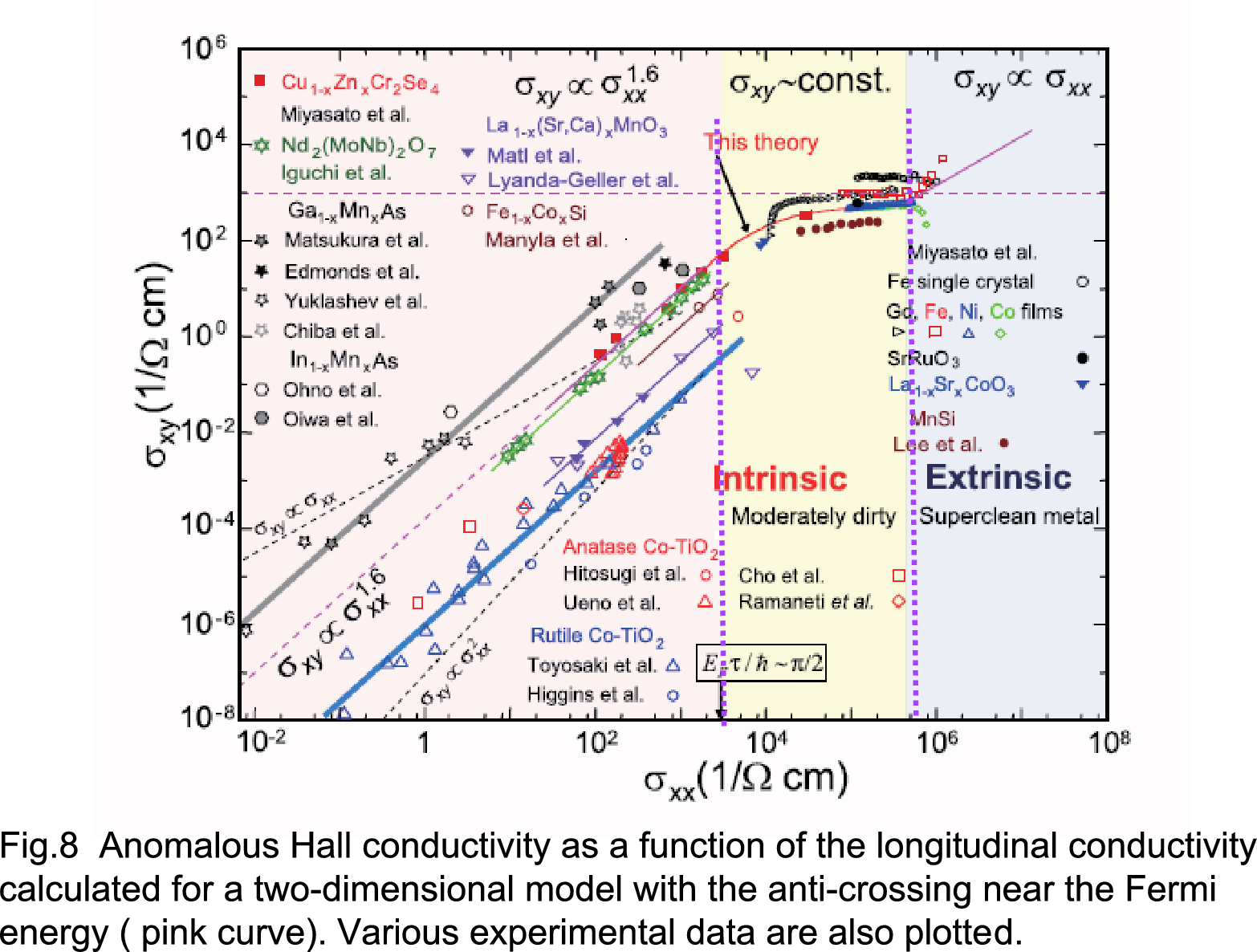}
\end{center}
\end{figure}

\clearpage

\begin{figure}[t]
\begin{center}
\includegraphics[width=16cm]{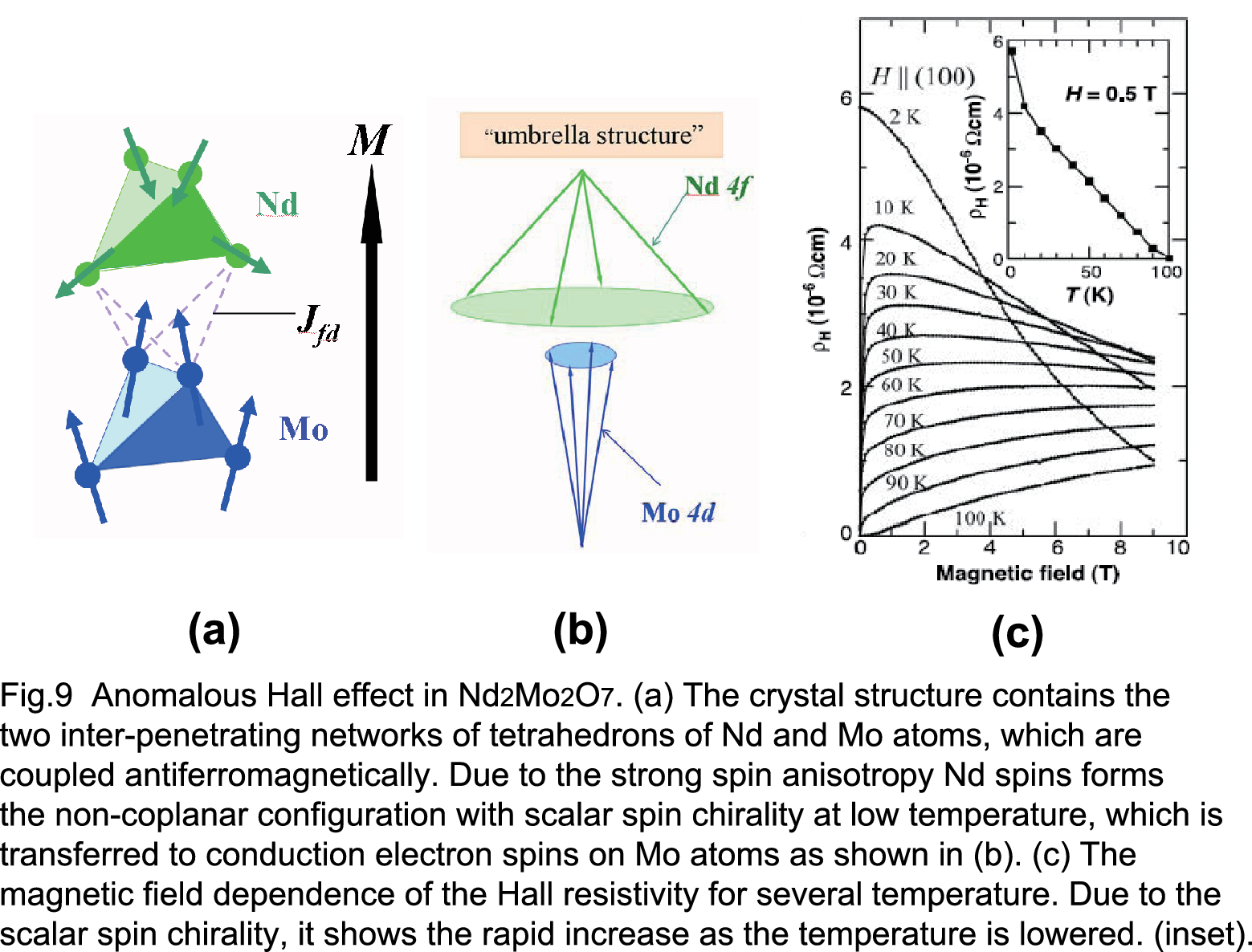}
\end{center}
\end{figure}

\clearpage

\begin{figure}[t]
\begin{center} 
\includegraphics[width=16cm]{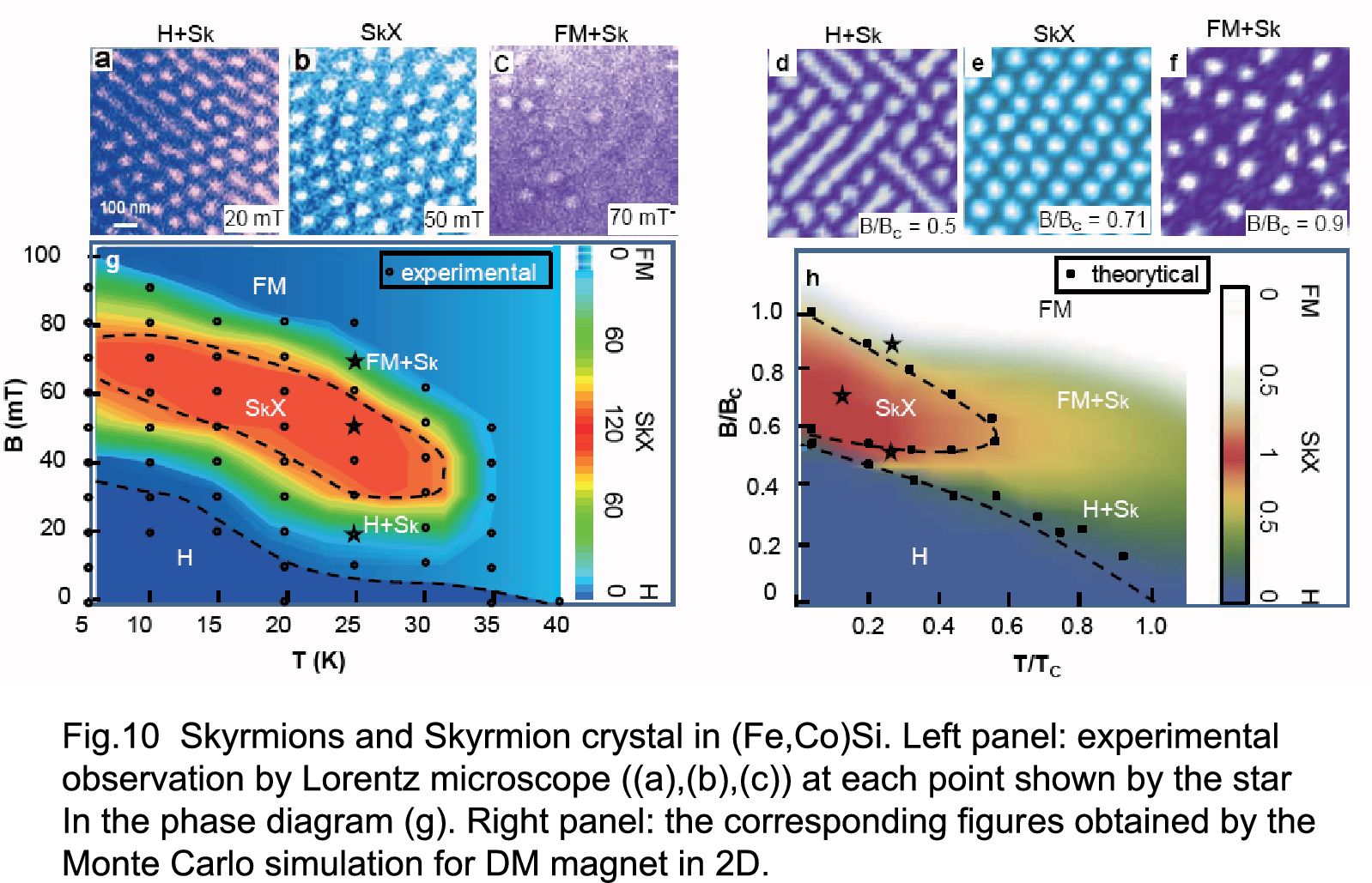}
\end{center}
\end{figure} 

\clearpage

\begin{figure}[t]
\begin{center}
\includegraphics[width=16cm]{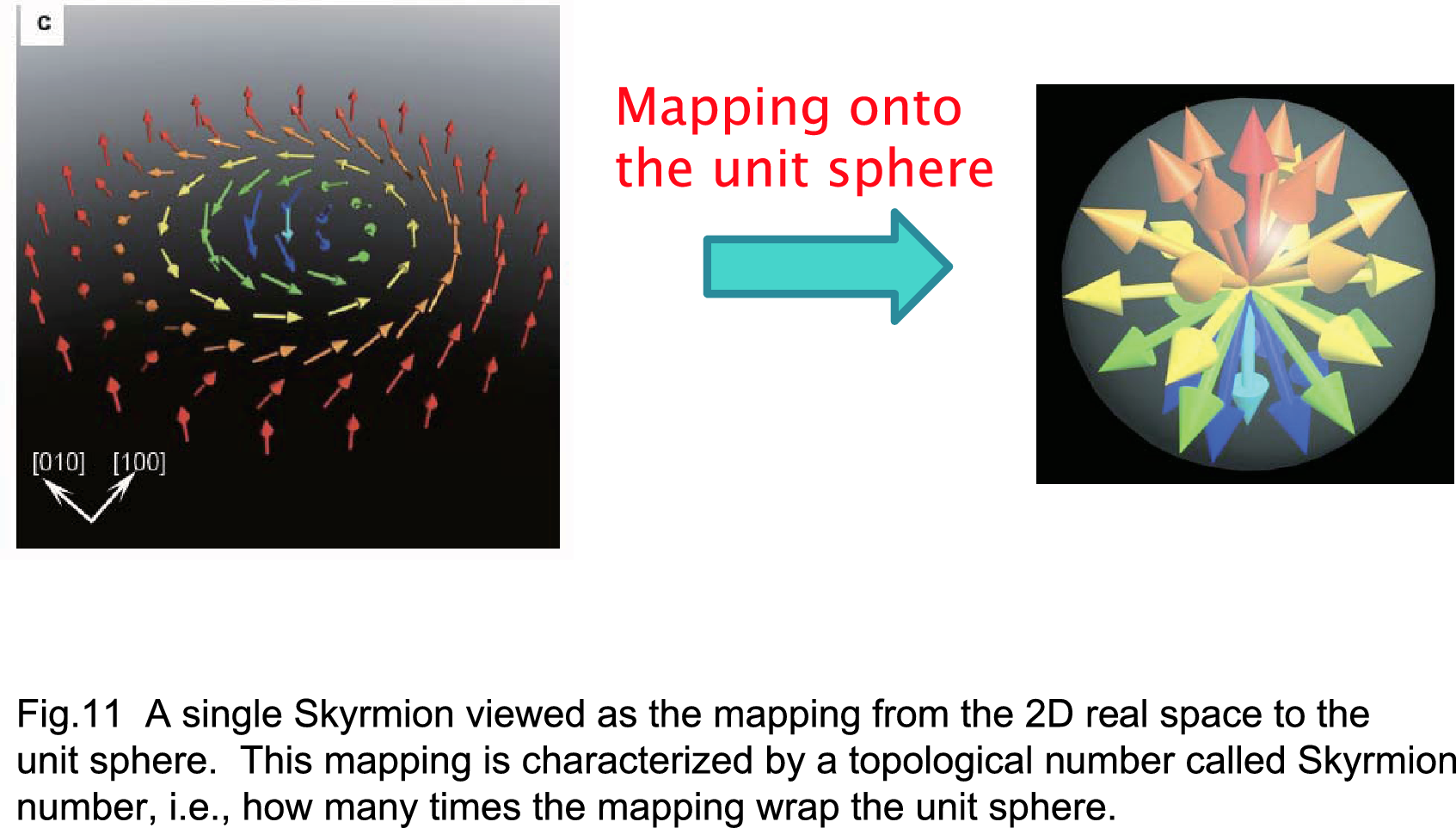}
\end{center}
\end{figure} 

\clearpage

\begin{figure}[t]
\begin{center}
\includegraphics[width=16cm]{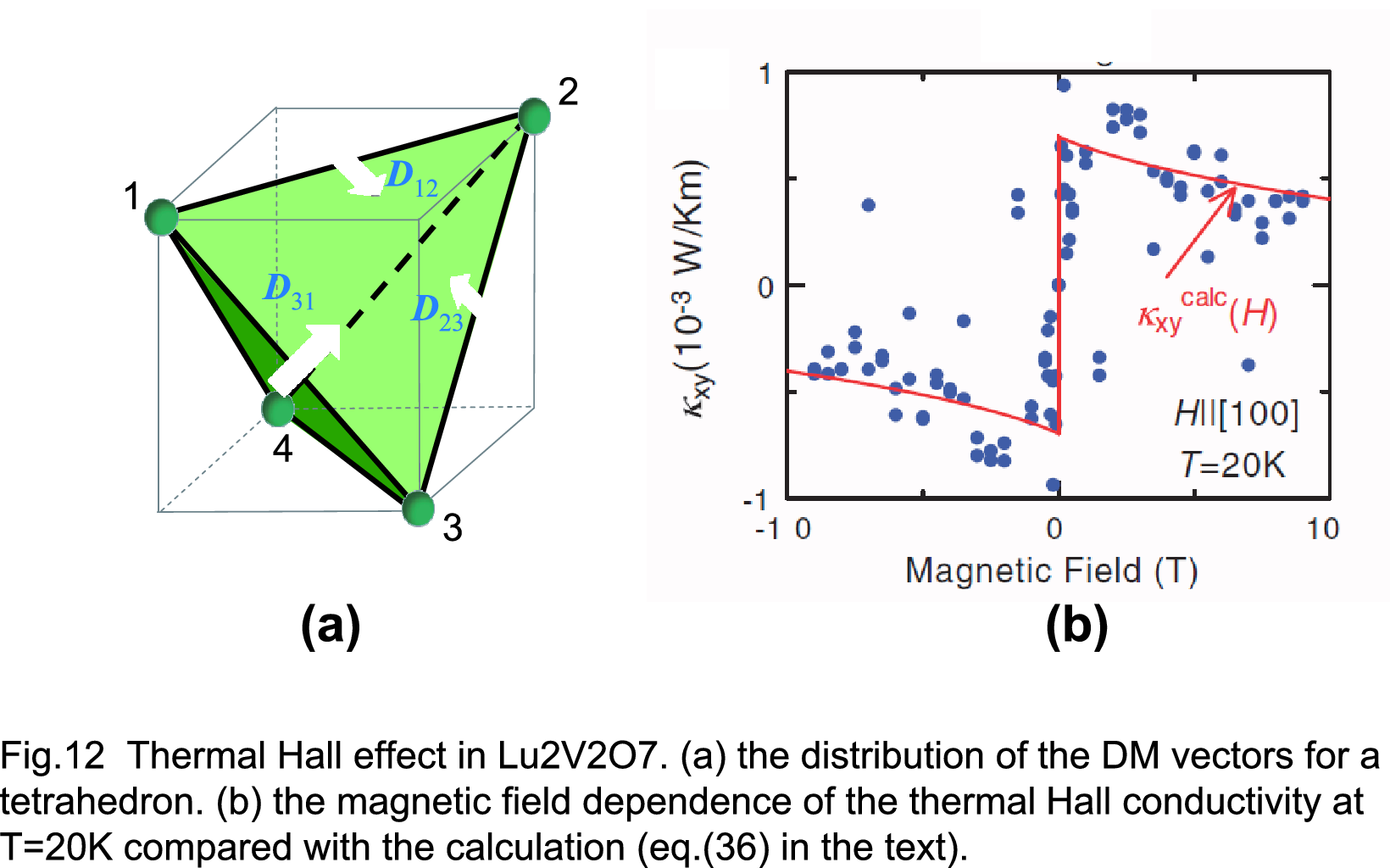}
\end{center}
\end{figure} 

\end{document}